\documentclass[aps,twocolumn,prx,floatfix,preprintnumbers, 10pt]{revtex4-2}
\usepackage{xpatch}
\makeatletter
\patchcmd{\@ssect@ltx}
    {\addcontentsline{toc}{#1}{\protect\numberline{}#8}}
    {}
    {}
    {}
\makeatother
\pdfoutput=1
\usepackage{color}
\usepackage{enumitem}

\makeatletter

\renewcommand{\thesection}{\arabic{section}}
\renewcommand{\thesubsection}{\thesection.\arabic{subsection}}
\renewcommand{\thesubsubsection}{\thesubsection.\arabic{subsubsection}}

\renewcommand{\p@section}{}
\renewcommand{\p@subsection}{}
\renewcommand{\p@subsubsection}{}

\let\oldappendix\appendix
\renewcommand{\appendix}{%
  \oldappendix
  \renewcommand{\thesection}{\Alph{section}}%
  \renewcommand{\thesubsection}{\thesection.\arabic{subsection}}%
  \renewcommand{\thesubsubsection}{\thesubsection.\arabic{subsubsection}}%
  \renewcommand{\p@section}{}%
  \renewcommand{\p@subsection}{}%
  \renewcommand{\p@subsubsection}{}%
  \renewcommand{\theHsection}{appendix.\Alph{section}}%
  \renewcommand{\theHsubsection}{\theHsection.\arabic{subsection}}%
  \renewcommand{\theHsubsubsection}{\theHsubsection.\arabic{subsubsection}}%
}

\makeatother

\usepackage{booktabs}
\usepackage{makecell}
\usepackage{cancel} 

\newcommand{\paragraf}[1]{\par\medskip\noindent{\bf #1}}

\usepackage{mathtools}
\usepackage[dvipsnames]{xcolor}
\usepackage{float}

\usepackage{hyperref}
\definecolor{amaranth}{rgb}{0.9, 0.17, 0.31}
\definecolor{forestgreen}{rgb}{0.13, 0.55, 0.13}
\definecolor{blue(munsell)}{HTML}{005567}
\definecolor{oxfordblue}{rgb}{0.0, 0.2, 0.4}
\definecolor{britishracinggreen}{rgb}{0.0, 0.26, 0.15}

\definecolor{SymTFTColores}{rgb}{0, 0.33, 70.1}

\definecolor{bblue}{rgb}{0.0, 0.58, 0.71}
\hypersetup{
pdfstartview={FitH}, % fits the width of the page to the window
pdftitle={}, % title
colorlinks=true, % false: boxed links; true: colored links
linkcolor=oxfordblue, % color of internal links (change box color with linkbordercolor)
citecolor=oxfordblue, % color of links to bibliography
filecolor=oxfordblue, % color of file links
urlcolor=oxfordblue% color of external links
}

\usepackage{pgfplots}
\pgfplotsset{compat=1.18}
\usepackage{amssymb}
\usepackage{amsfonts}
\usepackage{graphicx}
\usepackage{epstopdf}
\usepackage{dcolumn}
\usepackage{amsmath}
\usepackage{latexsym,bm}
\usepackage{amsthm}
\usepackage{slashed}
\usepackage{color}
\usepackage{url}
\usepackage{rotating}
\usepackage[normalem]{ulem}
\usepackage{faktor}
\usepackage{tikz-cd}
\usepackage{tensor}
\usepackage{array}
\usepackage{multirow}
\usepackage{comment}

\usepackage{tabularx}
\usepackage{makecell}
\usepackage[normalem]{ulem}
\usepackage{changepage}
 \numberwithin{equation}{section}

\definecolor{ssnblue}{HTML}{006B8F}

\newcommand{\SymTFTS}[3]{\Big( #1 \,\Big|\, #2 \,\Big|\, #3 \Big) }
\newcommand{\SymTFTEXTRA}[3]{\left( #1 \,\Big|\, #2 \,\Big|\, #3 \right) }

\newcommand{\SymTFTpic}[3]{%
\vcenter{\hbox{%
\begin{tikzpicture}[x=1pt,y=1pt]
  \fill[SymTFTColores, opacity=0.2] (0,0) rectangle (56,26);
  \draw[very thick] (0,0) -- (0,26);
  \draw[very thick] (56,0) -- (56,26);
  \node[font=\scriptsize] at (28,13) {$#2$};
  \node[font=\scriptsize, anchor=south, inner sep=1.5pt] at (0,26) {$#1$};
  \node[font=\scriptsize, anchor=south, inner sep=1.5pt] at (56,26) {$#3$};
\end{tikzpicture}}}}

\usepackage{makecell}
\newtheorem{theorem}{Theorem}[section]

\theoremstyle{definition}
\newtheorem{definition}[theorem]{Definition}
\theoremstyle{remark}

\theoremstyle{claim}

\usepackage{tikz}
\usepackage{diagbox}
\usetikzlibrary{positioning}
\usetikzlibrary{calc}
\usetikzlibrary{decorations.pathreplacing,calligraphy}

\usepackage{xstring}
\usetikzlibrary{decorations.pathmorphing} 
\usetikzlibrary{decorations.markings} 
\usetikzlibrary{arrows} 
\usetikzlibrary{shapes} 
\usetikzlibrary{matrix} 
\usetikzlibrary{positioning} 
\usepackage[english]{babel} 
\usepackage[autostyle]{csquotes}
\usepackage{pifont}
\usetikzlibrary{shapes.multipart}

\tikzset{->-/.style={decoration={
  markings,
  mark=at position .6 with {\arrow{Latex[length=1.5mm,width=1.5mm]}}},postaction={decorate}}}

\usepackage{quantikz}

\newcommand{\bea}{\begin{eqnarray}}
\newcommand{\eea}{\end{eqnarray}}
\newcommand{\be}{\begin{equation}}
\newcommand{\ee}{\end{equation}}
\newcommand{\ba}{\begin{aligned}}
\newcommand{\ea}{\end{aligned}}
\newcommand{\bit}{\begin{itemize}}
\newcommand{\eit}{\end{itemize}}
\newcommand{\ben}{\begin{enumerate}}
\newcommand{\een}{\end{enumerate}}

\newcommand{\id}{\text{id}}

\newcommand{\ad}{{\text{ad}}}

\newcommand{\Neu}{\text{Neu}}
\newcommand{\Dir}{\text{Dir}}

\newcommand{\PSU}{\text{PSU}}
\newcommand{\Bsym}{\mathfrak{B}^{\text{sym}}}
\newcommand{\Bphys}{\mathfrak{B}^{\text{phys}}}

\newcommand{\wh}{\widehat}

\newcommand{\half}{\tfrac{1}{2}}

\newcommand{\Z}{{\mathbb Z}}

\newcommand{\Q}{{\mathbb Q}}

\newcommand{\Fun}{\text{Fun}}

\newcommand{\Spin}{\text{Spin}}
\newcommand{\cA}{\mathcal{A}}
\newcommand{\cB}{\mathcal{B}}

\newcommand{\cE}{\mathcal{E}}

\newcommand{\cK}{\mathcal{K}}
\newcommand{\cL}{\mathcal{L}}
\newcommand{\cM}{\mathcal{M}}
\newcommand{\cN}{\mathcal{N}}

\newcommand{\cP}{\mathcal{P}}

\newcommand{\cS}{\mathcal{S}}
\newcommand{\cT}{\mathcal{T}}

\newcommand{\cX}{\mathcal{X}}

\newcommand{\cZ}{\mathcal{Z}}

\newcommand{\D}{\mathsf{D}}

\newcommand{\fB}{\mathfrak{B}}
\newcommand{\fZ}{\mathfrak{Z}}

\newcommand{\Hom}{\text{Hom}}

\renewcommand{\Vec}{\mathsf{Vec}}
\newcommand{\Rep}{\mathsf{Rep}}

\renewcommand{\dim}{\text{dim}}

\newcommand{\TwoVec}{2\mathsf{Vec}}
\newcommand{\TwoRep}{2\mathsf{Rep}}

\renewcommand{\Q}{\bm{Q}}

\makeatother

\makeatletter
\renewcommand\subsubsection{\@startsection{subsubsection}{3}{\z@}%
                                     {-3.25ex\@plus -1ex \@minus -.2ex}%
                                     {2.5ex \@plus .2ex}%
                                     {\centering\itshape\bfseries\small}}
\makeatother

\usepackage{physics}

\newcommand{\diag}{\text{diag}}

\newcommand{\Irr}{\text{Irr}}

\def\ad{\mathop{\mathrm{ad}}\nolimits}

\def\half{{\frac{1}{2}}}

\def\unit{{1\kern-.65ex {\rm l}}}
\def\1{{1\kern-.65ex {\rm l}}}

\usepackage{makecell}

\makeatletter
\newcommand\xlabel[2][]{\phantomsection\def\@currentlabelname{#1}\label{#2}}
\makeatother

\usepackage[status=draft,inline,nomargin]{fixme}
\fxusetheme{color}
\FXRegisterAuthor{ki}{aki}{\color{orange}KI}
\FXRegisterAuthor{ofw}{aofw}{\color{green!40!black}OFW}
\usepackage{adjustbox}

\usepackage{faktor}

\newcommand\restr[2]{{% we make the whole thing an ordinary symbol
  \left.\kern-\nulldelimiterspace % automatically resize the bar with \right
  #1 % the function
  \right|_{#2} % this is the delimiter
}}

\makeatletter
\def\l@subsubsection#1#2{}
\makeatother
\begin{document}

\title{Proliferation Transitions for Non-Abelian Anyons}

\author{
Sakura Sch\"afer-Nameki$^{1}$, 
Yunqin Zheng$^{2}$,
Andrea Antinucci$^{1}$ 
}

\affiliation{$^1$ Mathematical Institute, University of Oxford, Woodstock Road, Oxford, OX2 6GG, United Kingdom\\
$^2$ Kavli Institute for Theoretical Sciences,
University of Chinese Academy of Sciences, Beijing, 100190, China
}

\begin{abstract} 
\noindent 
We construct phase transitions that proliferate condensable anyons in general 2+1d topological orders, including non-abelian ones. The central tool that provides a systematic approach to this question is the Symmetry Topological Field Theory (SymTFT). For a given topological order, we  
identify the relevant symmetry from the transparent lines generated by the condensable anyons, and thereby realize the topological order in terms of a  3+1d SymTFT sandwich. The proliferation phase transition is realized by coupling scalar fields to the anyons purely on the symmetry boundary of the SymTFT. We illustrate the construction for abelian theories, as well as non-abelian ones, $D(S_3)$ and $SU(2)_k$ Chern-Simons theories, and extend it to anomalous anyons.

\end{abstract}

%%%%%%%%%%%%%%%%%%%%%%%%%%%
\maketitle
%%%%%%%%%%%%%%%%%%%%%%%%%%%%%%%%%%%%%

\tableofcontents

\section{Introduction}

Consider a (2+1)d Topological Quantum Field Theory (TQFT) $\cT$, i.e. a topological order (TO) described by a modular tensor category (MTC) of anyons \footnote{We refer to both the topological line and its endpoint as an anyon, when it does not lead to confusion. }. 
In addition, consider a collection of mutually local bosonic anyons with the structure that allows them to condense. Such a set of  anyons form a so-called {\bf condensable algebra} $\cA$ in $\cT$. 
Inserting a fine mesh of the anyons in $\cA$ proliferates these topological lines, and results in a new TQFT $\cT/\cA$. The map 
\be\label{TtoTA}
\cT \ \to \  \cT/\cA
\ee
is usually referred to as  ``condensing (the anyons in) the algebra $\cA$'', or gauging an anomaly-free (non-invertible) 1-form symmetry \cite{Cheng:2026qax,Moore:1989yh,lan2017hierarchy,Hsin:2018vcg}. This operation in itself is a purely topological manipulation, with  
$\cT$ and $\cT/\cA$ distinct topological phases of matter.

In this paper we answer the following question:
{\bf how can we realize \eqref{TtoTA} as a phase transition?} 
Since the gauging map \eqref{TtoTA} proliferates topological lines in $\cA$, a natural proposal is to couple the anyons $a\in\cA$ to scalar fields $\phi$ and condense them. We will refer to this as a Higgsing or {\bf  proliferation transition}.

For a single abelian anyon $a$ of order $n$ and spin $p/2n$ the question was answered in \cite{Cheng:2026qax}: a one-component scalar coupled to $a$ drives a transition whose Higgs phase is $(\cT\times U(1)_{-np})/\Z_n$, and this agrees with \eqref{TtoTA} precisely in presentations with $p=0$ \footnote{To be precise, one also need to deform the theory by monopole operators. See \cite{Cheng:2026qax} for details.}.  For non-abelian anyons--condensable algebras with constituents of quantum dimension larger than one, for which the 1-form symmetry generated by $\cA$ is non-invertible \footnote{Anyons are topological line operators, and the lines in $\cA$ generate a symmetry acting on the other anyons by braiding: a 1-form symmetry, non-invertible as soon as a constituent has quantum dimension larger than one. Condensability of $\cA$ is the statement that this symmetry is non-anomalous, i.e. gaugeable, with $\cT/\cA$ the gauged theory.} -- there are also a few examples \cite{Zhang:2024bye, Ji:2026yfj,Cordova:2025eim, Cordova:2025zkz, Lu:2026fid}, but no systematic {continuum} prescription is known so far.  
This is the question we will address in the present paper.

%%%%%%%%%%%%%%%%%%%%%%%%%%%%
\begin{table*}[t]
\centering
\footnotesize
\setlength{\arraycolsep}{4pt}
\adjustbox{max width=\textwidth}{$
\begin{array}{|c||c|c|c|}
\hline
\multicolumn{4}{|c|}{ }\\
\multicolumn{4}{|c|}{ 
\text{{\bf Input:} $\cA\subseteq\cT$ condensable algebra, \ $\langle\cA\rangle$ the generated sub-category, \
$\cE_\cA=\cZ_2(\langle\cA\rangle)\cong\Rep (G)$ its transparent lines} }\\
\multicolumn{4}{|c|}{ }\\
\hline\hline
&&&\\
 & \begin{array}{c} \text{\bf Type (i): Tannakian} \\ \langle\cA\rangle=\cE_\cA=\Rep (G) \end{array}
 & \begin{array}{c} \text{\bf Type (ii): Mixed} \\ \Vec\subsetneq\, \cE_\cA = \Rep (G) \, \subsetneq\langle\cA\rangle \end{array}
 & \begin{array}{c} \text{\bf Type (iii): Modular} \\ \cE_\cA=\Vec\,, \quad G=1 \end{array} \\
 &&& \\
\hline\hline
&&&\\
\begin{array}{c} \cM_\cA \\ \cN_\cA \end{array}
 & \begin{array}{c} \Vec \\ \cT/\Rep (G) \end{array}
 & \begin{array}{c} \langle\cA\rangle/\Rep (G) \\ \langle\cA\rangle'/\Rep (G) \end{array}
 & \begin{array}{c} \langle\cA\rangle \\ \langle\cA\rangle' \end{array} \\
 &&&\\
\hline
\cT
 & \begin{array}{c} \\ 
  \dfrac{\cN_\cA}{G^{(0)}} \\[8pt]
 %    \SymTFTS{\fB_{\Neu (G)}}{\fZ(G)}{\fB^{\cN_\cA}_{\Neu (G)}} \\[4pt]
     \SymTFTpic{\fB_{\Neu (G)}}{\fZ(G)}{\fB^{\cN_\cA}_{\Neu (G)}} \end{array}
 & \begin{array}{c} \\
 \dfrac{\cM_\cA\boxtimes\cN_\cA}{G^{(0)}} \\[8pt]
 %    \SymTFTS{\fB^{\cM_\cA}_{\Neu (G)}}{\fZ(G)}{\fB^{\cN_\cA}_{\Neu (G)}} \\[4pt]
     \SymTFTpic{\fB^{\cM_\cA}_{\Neu (G)}}{\fZ(G)}{\fB^{\cN_\cA}_{\Neu (G)}} \end{array}
 & \begin{array}{c} 
 \cM_\cA\boxtimes\cN_\cA \\[8pt]
     \text{SymTFT:} \ \ \fZ(1)=\TwoVec \end{array} \\ 
     & & & \\
\hline
\cT/\cA
 & \begin{array}{c} \\
 \dfrac{\cN_\cA}{H^{(0)}} \\[8pt]
 %    \SymTFTS{\fB_{\Neu (H)}}{\fZ(G)}{\fB^{\cN_\cA}_{\Neu (G)}} \\[4pt]
     \SymTFTpic{\fB_{\Neu (H)}}{\fZ(G)}{\fB^{\cN_\cA}_{\Neu (G)}} \end{array}
 & \begin{array}{c}\\
  \dfrac{\big(\cM_\cA/\wh\cA\big)\boxtimes\cN_\cA}{H^{(0)}} \\[8pt]
   %  \SymTFTS{\fB^{\cM_\cA/\wh\cA}_{\Neu (H)}}{\fZ(G)}{\fB^{\cN_\cA}_{\Neu (G)}} \\[4pt]
     \SymTFTpic{\fB^{\cM_\cA/\wh\cA}_{\Neu (H)}}{\fZ(G)}{\fB^{\cN_\cA}_{\Neu (G)}} \end{array}
 & \big(\cM_\cA/\cA\big)\boxtimes\cN_\cA \\
    &   & &\\
\hline
%%%%%
\begin{array}{c} 
\text{Proliferation} \\ \text{Transition}  \end{array}
 & \begin{array}{c} \\
     \SymTFTpic{\frac{\fB_{\Dir}\boxtimes \cS_{\cT,\cA}}{G^{(0)}}}{\fZ(G)}{\fB^{\cN_\cA}_{\Neu (G)}} \end{array}
 & \begin{array}{c}\\
     \SymTFTpic{\frac{\fB_{\Dir}\boxtimes \cS_{\cT,\cA}}{G^{(0)}}}{\fZ(G)}{\fB^{\cN_\cA}_{\Neu (G)}} \end{array}
 & S_{\cM_\cA\to\cM_\cA/\cA}\boxtimes\cN_\cA \\
    &   & &\\
\hline
%%%%%
\cS_{\cT, \cA} 
 & \begin{array}{c} \\
 S^{\text{LG}}_{H\subset G} \\
     \text{$G\to H$ Landau-Ginzburg$^*$ model} \end{array}
 & \begin{array}{c} \\
 S^{\text{LG}}_{H\subset G}\boxtimes S_{\cM_\cA\to\cM_\cA/\wh\cA} \\
     \text{Then repeat for $\wh\cA\subset\cM_\cA$} \end{array}
 & \begin{array}{c} \\
 S_{\cM_\cA\to\cM_\cA/\cA} \\
     \text{One scalar per generating anyon} \end{array} \\
     &&&\\
\hline
\end{array}
$}
\caption{The three types of condensable algebra $\cA$ in an MTC $\cT$, and the associated proliferation transitions. 
The algebra type hinges on the generated sub-category $\langle\cA\rangle$ and its transparent lines (the M\"uger center) $\cE_\cA=\cZ_2(\langle\cA\rangle)\cong\Rep (G)$: 
type (i) is when all lines in $\langle\cA\rangle$ are transparent, 
type (ii) when the transparent lines are  a proper sub-category, and type (iii) when there are no transparent lines, and $\langle \cA \rangle$ is modular. 
The key rows are $\cT$ and $\cT/\cA$ which provide SymTFT representations of the initial MTC and final MTC, as well as the proliferation transition theory and its SymTFT in the last two rows.
The data $\cM_\cA$ and $\cN_\cA$ are computed from $\cT, \cA$ and determine the BCs of the SymTFT; $\cS_{\cT,\cA}$ is the theory stacked on $\Bsym$ along the proliferation transition. 
Only the symmetry boundary (i.e. the left boundary) changes across the transition. 
Type (ii) is the general case: setting $\cM_\cA=\Vec$ gives (i), and setting $G=H=1$ gives (iii). Its transition is obtained by rerunning the construction on the strictly smaller pair $\wh\cA\subset\cM_\cA$, which terminates in type (i) or (iii). The constructions are in sections \ref{sec:typei} and \ref{sec:typeiii}. \label{tab:types}}
\end{table*}
%%%%%%%%%%%%%%%%%%%%%%%%%%%%

Anyon condensation and its relation to gapped boundaries, domain walls, and microscopic string-net Hamiltonians have been studied from complementary viewpoints \cite{Bais:2008ni,HungWanADE,ZhaoWanNonAbelian}. Related lattice constructions formulate condensation transitions in Landau–Ginzburg/Higgs language \cite{ZhaoWanLG}, while critical boundaries generated by competing condensates have been investigated in string-net SymTFT settings \cite{HungEtAlCFTFactory}. These approaches address microscopic realizations, boundary classifications, or particular critical models. The objective here is different: we seek a general continuum SymTFT organization which, for a condensable algebra $\cA\subset\cT$, determines the relevant symmetry from the M\"uger center $\cZ_2(\langle\cA\rangle)$ and realizes $ \cT\to\cT/\cA$ through dynamics confined to the symmetry boundary. Crucially, this provides a completely systematic approach to the problem.

We construct the proliferation transition for an arbitrary condensable algebra $\cA$ of any MTC $\cT$. We summarize the main result and construction of the transition in table \ref{tab:types}. 

The construction builds on the SymTFT realization of the MTC: $\cT$ can be presented in terms of a (3+1)d Symmetry Topological Field Theory (SymTFT)  \cite{Bhardwaj:2023ayw,Bhardwaj:2024qiv,Wen:2024qsg,Bhardwaj:2025piv} compactified on an interval with two topological boundary conditions (BCs). Since $\cT$ is topological, both are gapped BCs. 
However, crucially a SymTFT description has as input a symmetry of $\cT$.
We would like to associate a symmetry to a condensable algebra $\cA$, and construct $\cT$ as a SymTFT compactification for this symmetry. 

In particular, the anyons in $\cA$ should be placed on the symmetry boundary, $\Bsym$, whereas $\cT$ informs the physical boundary. 
However, the crux is in the complication that not every condensable algebra $\cA$ {corresponds to a symmetry}, more precisely, a condensable algebra is not a symmetry sub-category of $\cT$.

Our proposal is to associate to each algebra $\cA$ a symmetry $\cE_\cA$ which in turn fixes the SymTFT for $\cT$, which furnishes the  starting point of the proliferation transition: 
Given $\cA$, define first the minimal subcategory $\langle \cA\rangle$ which contains $\cA$. Then $\cE_\cA$ is the subcategory of transparent lines and thanks to powerful theorems in category theory, these are always of the type \cite{Deligne:2002cat} \footnote{In general the theorem by Deligne states that the category is either Tannakian $\Rep(G)$ or super-Tannakian $\cE_\cA=\Rep(G,z)$, where $z$ is a transparent fermion, namely it commutes with all the other lines of $\cE_\cA$ but has spin $\theta=-1$. We will in this paper not consider this option in detail.}
\be
\cE_\cA = \Rep (G) \,, 
\ee
i.e. the symmetry category generated by the irreducible representations (irreps) of a finite group $G$, with composition/fusion of irreps given by the tensor product of representations.  Note that for $G$ non-abelian this is a non-invertible symmetry (see \cite{Schafer-Nameki:2023jdn, Shao:2023gho} for reviews).  

The SymTFT associated to the pair $\cT$ and $\cA$ is then the 3+1d $G$ gauge theory (which has background fields for 0- and 1-form symmetries in 2+1d). The symmetry boundary $\Bsym$ is either a simple Neumann boundary condition $\Neu (G)$, which realizes purely the $\Rep (G)$ lines, or more generally we will see that the lines of $\cA$ are not contained in $\cE_\cA$ alone, but require stacking the boundary with a non-trivial TQFT $\cM_\cA$. Such {\bf non-minimal} or stacked boundary conditions are well-known for SymTFTs for 2+1d theories \cite{Bhardwaj:2023ayw,Bhardwaj:2024qiv,Wen:2024qsg,Bhardwaj:2025piv,Bhardwaj:2025jtf}.
The physical boundary is informed by the TQFT $\cT$.

The complete SymTFT construction for the pair $\cT$ and $\cA$ is then as follows:  compute the transparent lines inside the category generated by $\cA$, $\cE_\cA= \Rep (G)$. This determines the bulk to be the 3+1d $G$ gauge theory. 
The symmetry boundary $\Bsym$ is computed as follows. 
Define the MTC $\cM_\cA = \langle\cA\rangle/\Rep (G)$, which is $G$-symmetric. Take the standard Dirichlet boundary condition (Dir), which realizes the $G$ 0-form symmetry, and stack $\cM_\cA$ and gauge the diagonal $G$ 0-form symmetry
\be
\Bsym= \fB_{\Neu(G)}^{\cM_\cA}= {\fB_\Dir \boxtimes \cM_\cA\over G^{(0)}} \,.
\ee
For $\cM_\cA$ trivial, this is simply the Neumann BC  ($\Neu(G)$) for the $\Rep (G)$ 1-form symmetry. For non-trivial $\cM_\cA$ this supplies additional anyons on the boundary that are part of the symmetry. 
The physical boundary $\Bphys = \fB_{\Neu(G)}^{\cN_\cA}$ is similarly a non-minimal boundary stacked with $\cN_\cA = \langle \cA\rangle' /\Rep(G) $, where the $'$ indicates the centralizer in $\cT$. This allows us to rewrite $\cT$, in terms of a SymTFT sandwich as follows: 
\be 
\begin{split}
\begin{tikzpicture}
\begin{scope}[shift={(0,0)}]
\draw [SymTFTColores, fill= SymTFTColores, opacity = 0.2] 
(0,0) -- (0,2) -- (2,2) -- (2,0) -- (0,0) ; 
\draw [white] (0,0) -- (0,2) -- (2,2) -- (2,0) -- (0,0)  ; 
\draw [very thick] (0,0) -- (0,2) ;
\draw [very thick] (2,0) -- (2,2) ;
% \draw [ thick] (0,0.8) -- (2,0.8) ;
% \fill[] (0,0.8) circle (0.05cm);
% \fill[] (2,0.8) circle (0.05cm);
% \node[above] at (1,0.8) {$a$};
\node at  (1,1)  {$\fZ(G)$} ;
\node[above] at (-0.5,2) {$\Bsym=\fB_{\Neu(G)}^{\cM_\cA}$}; 
\node[above] at (2.2,2) {$\Bphys_\cT$}; 
\end{scope}
\begin{scope}[shift={(4,0)}]
\node at (-1,1) {$= \cT $} ;
\end{scope}
\end{tikzpicture}
\end{split}
\ee
Note that depending on $\cA$, both the bulk and BCs can vary, and give different presentations of $\cT$. 

Next we describe the proliferation transition in this SymTFT framework, which will happen solely on the symmetry boundary. 
Indeed, the lines of $\cA$ are contained in $\Bsym$ and 
gauging them, only changes the symmetry boundary. 
In order to realize the gauging map as a proliferation transition, it is natural to introduce a multiplet of boundary scalars $\phi=(\phi_a)$ on the symmetry boundary that couple to the  anyons $a$ in $\cA$. 
The transition is then realized purely on the symmetry boundary as follows: 
denoting the $\Bsym$ coupled to the scalars by $\Bsym_\phi$ we require that tuning the two signs of its mass term realizes the two phases:
\be \ba
\label{twophases}
m^2>0: \qquad &\Bsym_{\langle\phi\rangle =0}:\qquad  \cT\cr 
m^2<0: \qquad &\Bsym_{\langle\phi\rangle \not= 0}:\qquad  \cT/\cA\,.
\ea\ee
Tuning $m^2$ through zero is the proliferation transition. 
Note that this approach is very general, and applicable to abelian and non-abelian anyons alike;
the condensable algebras can be Lagrangian or non-maximal. 
We will illustrate this general principle with examples representing each of these situations.

\paragraf{Outline of the paper.} 
Setting up the SymTFT for a general pair $\cT$ and $\cA$ requires unpacking some mathematics, which we front-load in the paper as section \ref{sec:Algebras}. This mainly sets out the construction of $\cE_\cA$ and $\cM_\cA$. Each mathematical concept will have an important physics interpretation: $\cA$ defines a minimal sub-category in which it is contained. This will be denoted by $\langle \cA \rangle$. The transparent lines in this sub-category are precisely $\cE_\cA$ -- mathematically it is the M\"uger center of $\langle \cA\rangle$. We identify three types of algebras, which are characterized in terms of the relation of the transparent lines  $\cE_\cA$, to the generated subcategory: type (i) has $\cE_\cA = \langle\cA\rangle$, type (iii) has no transparent lines and is modular, and for type (ii) the transparent lines are a proper subset of lines in $\langle\cA\rangle$.

%%%%%%%%%%%%%%%%%%%%%%%%%%%%

{The main proposal of the paper is put forward in section \ref{sec:PTfromS}, and is summarized table \ref{tab:types}. 
Depending on the type of condensable algebra we can identify the proliferation transition for it: type (i) is of Landau-Ginzburg (LG) type (in fact it is an LG model for the gauged symmetry, $\Rep (G)$, and we refer to this as Landau-Ginzburg$*$),  type (iii) is modular and requires a case by case analysis, and 
type (ii) is in part  a  combination of an LG$*$ transition, and the case-by-case analysis of type (iii).

To put this proposal to work we first apply it to a general abelian anyon proliferation transitions in section \ref{sec:Zousuke}. This generalizes \cite{Cheng:2026qax} to arbitrary condensable algebras in abelian TOs. 

The main strength of the approach is that it applies to non-abelian TOs. We start in section \ref{sec:S3} with the first non-abelian example: proliferation transitions for any condensable algebra in $\cT= D(S_3)$. The condensable algebras are well-documented (see figure \ref{fig:S3Hasse}) \cite{Bais:2008ni, Bhardwaj:2024qrf}. Some of the algebras are simple subcategories of type (i), but here the main point is to show that we can construct a proliferation transition for {\bf all} condensable algebras of $D(S_3)$. 

In section \ref{sec:SU2k} we consider a  series of non-abelian MTCs $SU(2)_k$ and and the doubles $SU(2)_k\times SU(2)_{-k}$, which exhibit algebras of all three types, and we construct the associated proliferation transitions. 
}

In section \ref{sec:Ano} we go beyond condensable algebras, and ask what happens if we try to proliferate an arbitrary anyon $X\in \cT$, even if it is anomalous. From a TQFT viewpoint, while $X$ cannot be gauged, we can always stack a TQFT $\cX$ that cancels the anomaly and gauge the diagonal. We show that a slight modification of our general proposal can be applied to this case to canonically identify a minimal $\cX$ that allows for this generalized gauging to
\be \label{TXT}
    \frac{\cT\boxtimes \cX}{\cA_{\text{diag}}} \ .
\ee 
This provides a generalization to non-abelian anyons of the minimal theories for abelian anyon theories  in \cite{Hsin:2018vcg}. Our proposed SymTFT setup can, as in the non-anomalous case, be upgraded to a dynamical QFT by coupling the symmetry boundary with scalars, to obtain a field theory for the transition from $\cT$ to (\ref{TXT}).

We conclude with a summary and  outlook in section \ref{sec:Summary} and the appendices contain background on the SymTFT and some proofs.

\section{Condensable Algebras and their Subcategories}
\label{sec:Algebras}

Although the question underlying this paper is grounded firmly within physics,  to answer it we require a few insights from the mathematics of fusion categories. In particular condensable algebras, subcategories, the transparent lines (M\"uger center). This section will summarize concisely all the required data. Some of the math literature is 
\cite{Ostrikmodule, etingof2005fusion, EGNO, Muger:2003structure,  Kong:2014condensation}. Physics oriented readers may skip to the next section and refer back the definitions and theorems when needed. 

\begin{definition}[Condensable algebra]
\label{def:condensable-general}
Let $\cT$ be a unitary MTC, with braiding $c_{a,b}: a\otimes b \to b\otimes a$ and twist $\theta_a$.  
A {\bf condensable algebra} is an algebra object $(\cA,m,\eta)$, with multiplication $m:\cA\otimes\cA\to\cA$ and unit $\eta:\mathbf{1}\to\cA$, which is associative, commutative, connected, separable, and has trivial twist (i.e. is bosonic)
\be
\label{condensableconditions}
m\circ c_{\cA,\cA}=m\,,\qquad
\dim\Hom_{\cT}(\mathbf{1},\cA)=1\,,\qquad
\theta_{\cA}=\text{id}_{\cA}\,.
\ee
\end{definition}
Often a condensable algebra is just given in terms of its generating anyons (``simple objects") $\cA=\bigoplus_a n_a a$. However, this is in general not a complete description and inequivalent multiplications can exist on the same set of anyons, so-called twin algebras \cite{Warman:2026gfz}. We will focus on algebras that do not have this subtlety, and refer to the algebra in terms of its anyons. 

Condensing $\cA$ produces the topological order $\cT/\cA$, whose anyons are the local $\cA$-modules, of total dimension
\be
\label{quotientdimension}
\dim (\cT/\cA) = \frac{\dim\cT}{(\dim\cA)^2} \,.
\ee
Algebras with maximal dimension $\dim\cA=\sqrt{\dim\cT}$ are so-called  Lagrangian algebras with $\cT/\cA=\Vec$, and they correspond to gapped boundary conditions of $\cT$. Condensation is transitive: for condensable algebras $\cA\subseteq\cB$, with the inclusion a map of algebras, the {\bf residual algebra} $\cB/\cA$ is condensable in $\cT/\cA$ with
\be
\label{residualdimension}
\dim (\cB/\cA) = \frac{\dim\cB}{\dim\cA} \,,
\qquad
\frac{\cT/\cA}{\cB/\cA} \,\cong\, \frac{\cT}{\cB} \,,
\ee
see e.g. \cite{Davydov:2013structure}. This underlies all multi-step transitions in this paper. 

Although condensable algebras come with a multiplicative structure as an algebra, this does not mean that an algebra closes under the fusion -- i.e. the composition of anyons $a\otimes b$ in the MTC $\cT$. In fact most algebras do not close and thus do not form sub-symmetries (or sub-categories) themselves. However we can define the minimal sub-category, that contains $\cA$: 

\begin{definition}[Generated sub-category]
\label{def:generated}
For a condensable algebra $\cA$ with support $\mathrm{supp}(\cA)=\{a\ \text{simple}:\ \Hom(a,\cA)\neq0\}$, the {\bf generated sub-category} $\langle\cA\rangle\subseteq\cT$ is the smallest fusion sub-category containing $\mathrm{supp}(\cA)$.
\end{definition}

A fusion sub-category inherits braiding and twist from $\cT$, but the braiding is not necessarily non-degenerate, and $\langle\cA\rangle$ is in general pre-modular. A way to detect this is the M\"uger center, which are all the transparent lines: 

\begin{definition}[M\"uger center and Centralizer]
\label{def:Muger-center-frame}
For a braided fusion sub-category $\cB\subseteq\cT$, the M\"uger center is
\be
\label{eq:Muger-center-BA}
\cZ_2(\cB) :=\left\{x\in\cB \, :\, c_{y,x}c_{x,y}=
\id_{x\otimes y}\ \forall\, y\in\cB\right\}\,,
\ee
whose objects are called {\bf transparent}. 
The centralizer $\cB'$ is similarly defined but now within $\cT$:
\be
\cB' := \left\{x\in\cT \, :\, c_{y,x}c_{x,y}=
\id_{x\otimes y}\ \forall\, y\in\cB\right\}\,.
\ee
\end{definition}

The following result will play a crucial role in our construction:
\begin{theorem}[M\"uger \cite{Muger:2003structure} (Theorem 3.2, Corollary 3.3 and Theorem 4.2)]
\label{thm:Muger-gluing}
Let $\cB\subseteq\cT$ be a braided fusion sub-category of the MTC $\cT$. Then
\be
\label{eq:BA-centralizer-dimension}
\dim (\cB)\,\dim (\cB')=\dim(\cT)\,,
\qquad
\cB''=\cB\,.
\ee
Moreover $\cB\cap\cB'=\cZ_2(\cB)$ by definition, so that in particular $\cZ_2(\cB')=\cZ_2(\cB)$.
If $\cZ_2(\cB)=\Vec$, then $\cT$ factorizes,
\be
\label{eq:Muger-factorization-BA}
\cB\boxtimes\cB' \cong \cT\,.
\ee
\end{theorem}

A pre-modular category with symmetric braiding and trivial twists, namely $\cB\equiv \cZ_2(\cB)$, is called {\bf Tannakian}. An important theorem by Deligne \cite{Deligne:2002cat} states that Tannakian categories are equivalent to $\Rep (G)$ for a finite group $G$ (again with the disclaimer about the case of transparent fermions). In particular the M\"uger center of any subcategory is of the form $\Rep (G)$.

We will denote the M\"uger center of the generated category of $\cA$ as follows: 
\be 
\cE_\cA:=\cZ_2(\langle\cA\rangle) \cong \Rep (G) \,.
\ee 
This is the crucial piece of data that will allow us to construct the SymTFT. 
 There are essentially three types of condensable algebras that will be useful to separate in the following.
\paragraf{Types of Condensable Algebras:}
Let $\cA$ be a condensable algebra in an MTC $\cT$, $\langle\cA\rangle$ its generated sub-category that is generically pre-modular and $\cE_\cA$ the M\"uger center of transparent lines of $\langle\cA\rangle$. 
Then one of the following three options occurs (tautologically, but the separation into these cases is useful from a physical perspective)
\begin{enumerate}[label=(\roman*),leftmargin=2em]
\item $\langle\cA\rangle=\cE_{\cA}\cong\Rep(G)$ for a finite group $G$.
\item $\Vec\subsetneq\cE_{\cA}\subsetneq\langle\cA\rangle$.
\item $\cE_{\cA}=\Vec$, so that $\langle\cA\rangle$ is modular.
\end{enumerate}

For type (i) algebras, $\langle \cA\rangle$ is Tannakian, i.e. $\Rep (G)$, and we can carry out the Higgsing completely on a $\Neu (G)$ BC of the SymTFT. If there are transparent fermions, in which case the category is in fact super-Tannakian $\Rep(G,z)$, we will discuss this whenever it occurs but will not consider this in our general discussion for clarity of exposition. 

For type (iii) algebras, the M\"uger center is trivial and the category is modular. Type (ii) is an admixture of type (i) and type (iii) condensations. 
In the (iii) case $\langle\cA\rangle$ is modular and we can apply theorem \ref{thm:Muger-gluing}, which implies that $\cT$ in fact factorizes.

This concludes the summary of the main mathematical concepts. We now turn to the description of the physical setup and proposal.

\section{Proliferation Transitions from the SymTFT}
\label{sec:PTfromS}

Our starting point is an MTC $\cT$ and a condensable algebra $\cA$ in $\cT$. The first step is to realize $\cT$ as a SymTFT for a suitable symmetry that then allows proliferation of the anyons in $\cA$. 
We give a quick crashcourse on the SymTFT focusing on 2+1d theories, in appendix \ref{app:symtft}. 

In section \ref{sec:gum} we give a more physical description of the types of algebras introduced in section \ref{sec:Algebras}. 
For pedagogical reasons we will then first discuss the case when $\cA$, or its generated sub-category $\langle \cA \rangle$, are of $\Rep (G)$ type, (type (i) algebras in the nomenclature of section \ref{sec:Algebras}), and then consider the general case subsequently.  

\subsection{SymTFT Realization and Algebra Types}
\label{sec:gum}

Central to the construction is to identify the relevant symmetry structure associated to $\cA$, and to then realize $\cT$ and $\cT/\cA$ in terms of SymTFT compactifications 
\be
\label{framesandwich}
\underbrace{\SymTFTS{\Bsym}{\fZ(G)}{\Bphys}}_{=\cT} \stackrel{\langle \phi \rangle}{\longrightarrow}
\underbrace{\SymTFTS{{\Bsym}'}{\fZ(G)}{\Bphys}}_{=\cT/\cA} \,.
\ee
The main non-trivial point here is to associate to any condensable algebra $\cA$ a symmetry, which is not just $\cA$ (in fact $\cA$ itself often is not a symmetry at all): define first the fusion sub-category $\langle\cA\rangle\subseteq\cT$ that is {\bf generated} by $\cA$, i.e. the minimal category which is the closure under fusion. 
The SymTFT is constructed from the {\bf transparent lines} 
in $\langle\cA\rangle$, i.e. those braiding trivially with all the lines in $\langle\cA\rangle$, given by the {\bf M\"uger center $\cE_\cA$}. The key symmetry input is the fact that the transparent lines form the representation category of a finite group $G$
$\Rep (G)$ \footnote{This assumes we only have transparent bosont, whereas there is also the possibility of a transparent fermion, in which this is $\Rep (G,z)$.} and the SymTFT is simply the 3+1d $G$-gauge theory $\mathfrak{Z}(G)$. 

The structure of the transparent lines informs entirely the SymTFT realization, and the resulting proliferation transition: 
Exactly one of the following situations holds -- these are the {\bf types of condensable algebras} in section \ref{sec:Algebras}, which we now recast in terms of their physical properties:
\begin{itemize}
\item[(i)] All lines in $\langle \cA\rangle$ are transparent and $\langle\cA\rangle=\Rep (G)$ for some finite group $G$. 
Mathematically this means $\langle\cA\rangle$ is its own M\"uger center:  $\langle\cA\rangle = \cE_\cA$. 
\item[(ii)] The transparent lines of $\langle\cA\rangle$ form a proper non-trivial sub-symmetry of $\langle \cA \rangle$. 
\item[(iii)] $\langle\cA\rangle$ has no transparent lines, and $\langle\cA\rangle$ is modular and $\cT$ has a factor $\langle\cA\rangle$. \end{itemize}
We will explain now each of these three instances and provide examples for them: type (i) in section \ref{sec:typei} and type (ii) and (iii) in section \ref{sec:typeiii}.

\subsection{Proliferating $\Rep (G)$} 
\label{sec:typei}

The type (i) algebras are the simplest type for our purposes, where   $\cA$ itself is a sub-category of $\cT$, in which case it is also $\Rep (G)$ for some finite group $G$. 
Examples are the pure charge algebras in quantum doubles of group $D(G)$, which form $\cA=\Rep (G)$, or $\cA= \Rep (G/N)$ for $N$ a normal subgroup of $G$. 
However, even if $\cA$ itself is not a sub-category, its generated sub-category $\langle\cA\rangle$ may be $\Rep (G)$ and all its lines are transparent. We treat these types in a single setup -- they will only differ in the subset of anyons that will end in Higgs fields to trigger the transition. 
These are all type (i) or Tannakian algebras (as the relevant symmetries are Tannakian $\Rep(G)$ subcategories of $\cT$). 

\paragraf{From an Algebra $\cA$ to a Symmetry.}
A type (i) algebra is always a coset algebra. This follows from  \cite{lan2017modular} (Proposition 3.6, see also example 4.11 in \cite{davydov2010witt}), which discusses the algebras inside $\Rep (G)$: these are labeled by subgroups $H \leq G$  and are \footnote{Note that this fixes the subgroup $H$ only up to conjugation.}
\be
\label{cosetalgebra}
\cA=\Fun(G/H)\,,
\ee
the functions $G/H\to \mathbb{C}$ with point-wise multiplication. This is the algebra of functions on the coset space $G/H$. Only if $G/H$ is a group, i.e. if $H$ is normal we get a $\Rep (G/H)$ subcategory. 
Alternatively it is the induced representations from $H$ to $G$: for an irrep $R$ of $H$ we get a representation of $G$ by 
$\text{Ind}_H^G R=\mathbb{C}[G] \otimes_{\mathbb{C}[H]} R$. The functions on the coset are then simply the induced representation of the trivial irrep $1$, which can be subsequently decomposed into irreps of $G$ as follows: 
\be \label{cosetindu}
\cA= \text{Ind}_H^G  \, 1 
= \bigoplus_{R\in\Irr (G)}\dim\big(R^H\big)\,R\,,
\ee
where $R^H$ is the subspace of $H$-fixed vectors in the irrep $R$. For $H=1$, $\cA=\Rep (G)$ -- which is the case when $\cA$ is a subcategory. 

A technical, though quite important point is the following: 
In fact the generated subcategory for $\cA= \Fun(G/H)$ is 
\be
\langle \cA\rangle=\Rep(G / \text{core}_G(H))\,,
\ee
where the core, defined as 
\be 
\text{core}_G(H) =\bigcap_{g\in G}gHg^{-1}\,,
\ee 
is the largest normal subgroup of $G$ contained in $H$. 
In the following we will always consider $G$ and $H$, such that $H$ in $G$ is corefree, i.e. $\text{core}_G(H)=1$. This simply means, we pick the minimal $G$ to represent $\cA$. That way 
\be
\langle \cA \rangle = \Rep (G) \,.
\ee
We will exemplify this with $D(S_3)$ later. 

\paragraf{SymTFT Sandwich of $\cT$.}
We now construct the SymTFT sandwich for $\cT$.  The bulk SymTFT is the (3+1)d $G$ gauge theory $\fZ (G)$, whose topological defects are the flux surfaces $\Q_2^{[g]}$ and Wilson lines $\Q_1^{\bm R}$ (appendix~\ref{app:symtft}). The symmetry boundary is chosen as $\Neu(G)$, i.e. is Neumann for $G$. Here  $\Q_2^{[g]}$ end, $\Q_1^{\bm{R}}$ generate the 1-form symmetry $\Rep(G)$:
\be
\label{symmetrytrade}
\ba
\fB_{\Neu(G)}: &\quad \Q_2^{[g]}\ \text{end}\,,\quad \Rep(G)\ \text{realized by } Q_1^{\bm{R}} \,.\cr
\ea
\ee
The fully gauged, dual, boundary condition, Dirichlet, realizes the $G^{(0)}$ 0-form symmetry:
\be
\ba
\fB_\Dir:& \quad  \Q_1\ \text{end}\,,\quad G^{(0)}\ \text{realized by }\Q_2 \,.
\ea
\ee
The physical boundary is obtained from a non-minimal BC, which is obtained from the Dirichlet BC $\fB_{\text{Dir}}$ by stacking with a $G^{(0)}$-graded TQFT, 
\be 
\cN=\frac{\cT}{\Rep (G)} \,,
\ee
and gauging the diagonal $G^{(0)}$: 
\be\label{nonmini}
\fB_{\Neu(G)}^\cN = {\fB_{\Dir} \boxtimes \cN  \over G^{(0)} } \,.
\ee
We can write the original theory as the SymTFT sandwich
\be
\label{TSymTFT}
\cT=\SymTFTS{\fB_{\Neu (G)}}{\fZ(G)}{\fB^{\cN}_{\Neu (G)}}
\,.
\ee

\paragraf{SymTFT Sandwich of $\cT/\cA$.}
We now construct the SymTFT sandwich for $\cT/\cA$. Both the bulk SymTFT and the physical boundary are unchanged, being $\fZ(G)$ and $\fB^{\cN}_{\Neu (G)}$ respectively. Gauging $\cA$ only changes the symmetry boundary condition from $\fB_{\Neu (G)}$ to $\fB_{\Neu (H)}$, where 
\begin{equation}\label{eq:NeuH}
    \fB_{\Neu(H)} = \frac{\fB_{\Dir}}{H^{(0)}} = \frac{\fB_{\Dir}\boxtimes Q_{G/H}}{G^{(0)}}\,.
\end{equation}
The topological lines supported on $\fB_{\Neu (H)}$ form $\Rep(H)$. $Q_{G/H}$ is the 3d TQFT with $|G/H|$ number of vacua, and each vacuum is trivially gapped. We can write the theory $\cT/\cA$ as the SymTFT 
\begin{equation}
    \cT/\cA=\SymTFTS{\fB_{\Neu (H)}}{\fZ(G)}{\fB^{\cN}_{\Neu (G)}}
\,.
\end{equation}

\paragraf{Proliferation Transition.}
To realize the phase transition $\cT\to \cT/\cA$, we only need to realize the phase transition on the symmetry boundary $\fB_{\Neu (G)}\to \fB_{\Neu (H)}$. This can be achieved as follows. 

We start with the symmetry boundary being the $\fB_{\Dir}$, which has $G^{(0)}$ symmetry. Now introduce the Landau-Ginzburg (LG) theory triggered by scalar with a Higgs potential, which is denote as $S^{\text{LG}}_{H\subset G}$. Now we gauge the diagonal $G^{(0)}$ that acts both on the $\fB_{\Dir}$ and the LG theory
\begin{equation}\label{eq:tranSym}
    \frac{\fB_{\Dir}\boxtimes S^{\text{LG}}_{H\subset G}}{G^{(0)}}\,.
\end{equation}
In the $G$ preserving phase, $ S^{\text{LG}}_{H\subset G}$ is trivially gapped, hence the symmetry boundary is $\fB_{\Neu(G)} = \fB_{\Dir}/G^{(0)}$. In the phase where $G$ is spontaneously broken to $H$, $ S^{\text{LG}}_{H\subset G}= Q_{G/H}$, and  \eqref{eq:tranSym} is precisely $\fB_{\Neu(H)}$, as \eqref{eq:NeuH}. In short, the proliferation phase transition theory written as a SymTFT is 
\be\label{eq:LG gauged}
    \cP_{\cT\to \cT/\cA}= 
    \SymTFTEXTRA{\frac{\fB_{\Dir}\boxtimes S^{\text{LG}}_{H\subset G}}{G^{(0)}}}{\fZ(G)}{\fB_{\Neu(G)}^{\cN}}\,.
\ee
The universality class of the transition coincides with that of the $G\to H$ LG$^*$ theory \footnote{Following the standard convention, the star means $G$ is gauged. For instance, Ising$^*$ universality class is related to Ising university class by gauging $\Z_2$. }.

\subsection{General Type Algebras}
\label{sec:typeiii}

We now discuss the general case for any condensable algebra $\cA$, which covers in particular also type  (ii) and modular  type (iii) algebras. For this we need to extend the framework slightly. 
Again let $\langle\cA\rangle$ be the generated subcategory of $\cA$, with M\"uger center (again with the restriction to the transparent boson case)
\be 
\cE_\cA= \Rep (G) \,.
\ee 
We define two MTCs starting from this data, which will be the input into the SymTFT boundary conditions:
\be
\label{MXFrames}
\cM_\cA = \frac{\langle\cA\rangle}{\Rep (G)} \,,\qquad
\cN_\cA= \frac{\langle\cA\rangle'}{\Rep (G)}\,,
\ee
where $\langle\cA\rangle'$ is the centralizer of $\langle\cA\rangle$ in $\cT$, i.e. all lines of $\cT$ braiding trivially with it (Definition \ref{def:Muger-center-frame}). The quotients condense the transparent bosons, so that $\cM_\cA$ and $\cN_\cA$ are modular \cite{Mueger:1999yb}.
Also, because we gauged the symmetry $\Rep (G)$, the resulting MTCs have a dual $G^{(0)}$ 0-form symmetry. 
$\cT$ can be factorized as 
\be\label{eq:cTfactorize}
    \cT = \frac{\cM_\cA \boxtimes \cN_\cA}{G^{(0)}}\,.
\ee
As a special case, when $\langle \cA \rangle = \Rep(G)$, $\langle \cA \rangle' = \cT$, and \eqref{eq:cTfactorize} reduces to $\cT= \frac{\cT}{\Rep(G)}/G^{(0)}$ which is true. 

To show \eqref{eq:cTfactorize}, we first note that Theorem \ref{thm:Muger-gluing} applied to the generated category $\langle \cA \rangle$ implies 
\be
\label{frameoverlap}
\langle\cA\rangle\cap\langle\cA\rangle'=\Rep (G)\,,\qquad
\dim\langle\cA\rangle\,\dim\langle\cA\rangle'=\dim\cT\,,
\ee
i.e. $\langle \cA \rangle$ and its centralizer intersect in the transparent lines. The two quotients by the intersection \eqref{MXFrames} are modular of dimensions $\dim\langle\cA\rangle/|G|$ and $\dim\langle\cA\rangle'/|G|$, whereas  $\dim\big(\cT/\Rep (G)\big)=\dim\cT/|G|^2$. The quotients $\cM_\cA$ and $\cN_\cA$  centralize each other and their dimensions multiply to $\dim(\cT/\Rep (G))$, so each is the centralizer of the other in $\cT/\Rep(G)$. We can now apply Theorem \ref{thm:Muger-gluing} inside $\cT/\Rep (G)$ to get 
\be
\label{modularizedfactorization1}
\frac{\cT}{\Rep (G)} = \cM_\cA\boxtimes\cN_\cA\,.
\ee
$\cT$ is recovered by gauging the dual $G^{(0)}$ of \eqref{modularizedfactorization1}, which is \eqref{eq:cTfactorize}.

\paragraf{SymTFT Sandwich of $\cT$.} We now construct the SymTFT sandwich for $\cT$. Since our purpose is to extract the topological lines contained in $\cA$, the category of lines living on the symmetry boundary is $\langle \cA \rangle$. Moreover, since the M\"uger center $\cE_\cA=\Rep(G)$ braids trivially with all lines in $\cA$, they can be lifted to the bulk SymTFT. Therefore, the SymTFT is the (3+1)d $G$ gauge theory $\fZ(G)$. The symmetry boundary is
\be \label{Baba}
    \Bsym=\fB^{\cM_\cA}_{\Neu (G)} = \frac{\fB_{\Dir} \boxtimes \cM_\cA}{G^{(0)}}\,.
\ee
The physical boundary is
\be\label{Papa}
    \Bphys=\fB^{\cN_\cA}_{\Neu (G)} = \frac{\fB_{\Dir} \boxtimes \cN_\cA}{G^{(0)}}\,.
\ee
Indeed, it can be verified that 
\begin{equation}\label{eq:Tgeneral}
    \cT= \SymTFTS{\fB^{\cM_\cA}_{\Neu (G)}}{\fZ(G)}{\fB^{\cN_\cA}_{\Neu (G)}}\,.
\end{equation}
To show \eqref{eq:Tgeneral}, we first rewrite $\cN_\cA$ as a SymTFT sandwich with $G^{(0)}$ symmetry
\begin{equation}
    \cN_\cA= \SymTFTS{\fB_{\Dir}}{\fZ(G)}{\fB^{\cN_\cA}_{\Neu (G)}}\,.
\end{equation}
Tensoring both sides by $\cM_\cA$, and gauging the diagonal $G^{(0)}$ symmetry precisely gives rise to \eqref{eq:Tgeneral}. 

\paragraf{SymTFT Sandwich of $\cT/\cA$. }
We now construct the SymTFT sandwich for $\cT/\cA$. Both the bulk SymTFT and the physical boundary are unchanged, being $\fZ(G)$ and $\fB_{\Neu(G)}^{\cN_\cA}$ respectively. Gauging $\cA$ only changes the symmetry boundary condition $\fB_{\Neu(G)}^{\cM_\cA}$.

To see what it changes to, we first note that the transparent anyons in $\cA$ also form an algebra 
\begin{equation}
    \label{transparentpart}
\cA \cap \cE_\cA = \Fun(G/H) \,,\qquad H\leq G \,.
\end{equation}
Gauging $\Fun(G/H)$ on the symmetry boundary yields the boundary condition 
\begin{equation}\label{eq:symbdy1}
    \frac{\fB_{\Dir}\boxtimes \cM_{\cA}}{H^{(0)}} = \frac{\fB_{\Dir}\boxtimes \cM_{\cA} \boxtimes Q_{G/H}}{G^{(0)}}\,,
\end{equation}
where $Q_{G/H}$ is defined in (\ref{eq:NeuH}).
Substituting it to the SymTFT sandwich yields the theory 
\begin{equation}
    \frac{\cT}{\Fun(G/H)}= \frac{\cM_\cA\boxtimes \cN_\cA}{H^{(0)}} \,.
\end{equation}
To achieve gauging $\cA$, we need to further gauge an algebra $\widehat\cA$--the image of $\cA$ under the condensation map 
\begin{equation}
    F: \langle \cA\rangle \to \cM_\cA\,,
\end{equation}
i.e. $\widehat\cA=F(\cA)$, with $\dim \wh{\cA} =\dim\cA/[G:H]$. The symmetry boundary \eqref{eq:symbdy1} reduces to 
\begin{equation}
    \frac{\fB_{\Dir}\boxtimes \cM_{\cA}/\widehat{\cA}}{H^{(0)}} = \frac{\fB_{\Dir}\boxtimes \cM_{\cA}/\widehat{\cA} \boxtimes Q_{G/H}}{G^{(0)}}\,.
\end{equation}
Substituting it to the SymTFT sandwich yields the final gauged theory
\be
\label{masterendpoint}
\cT/\cA= \frac{\big(\cM_\cA/\wh\cA\big) \boxtimes \cN_\cA}{H^{(0)}} 
\ee
with $\dim \wh{\cA} =\dim\cA/[G:H]$.
We prove this in appendix \ref{app:MasterEq}. This splits -- formally --  the proliferation of the anyons in $\cA$ up into two steps: 
\be
\label{stagedchain}
\cT\ \xrightarrow{\ \Fun(G/H)\ }\ \cT_1= \frac{\cM_\cA\boxtimes\cN_\cA}{H^{(0)}}\ \xrightarrow{\ \ \wh\cA\ \ }\ \cT/\cA\,.
\ee
In summary, the SymTFT sandwich for the gauged theory $\cT/\cA$ is 
\begin{equation}\label{eq:T/Ageneral}
\boxed{
    \cT/\cA= \SymTFTEXTRA{\frac{\fB_{\Dir}\boxtimes \cM_{\cA}/\widehat{\cA} \boxtimes Q_{G/H}}{G^{(0)}}}{\fZ(G)}{\fB^{\cN_\cA}_{\Neu (G)}}} \,.
\end{equation}

\paragraf{Proliferation Transition. }
To realize the phase transition $\cT\to \cT/\cA$, we only need to realize the phase transition on the symmetry boundary 
\begin{equation}
    \fB_{\Neu(G)}^{\cM_\cA} \longrightarrow \frac{\fB_{\Dir}\boxtimes \cM_{\cA}/\widehat{\cA} \boxtimes Q_{G/H}}{G^{(0)}} \,.
\end{equation}
It is straightforward to see that this phase transition can be achieved by replacing the symmetry boundary to the following boundary condition 
\be 
\Bsym \ \mapsto\  \frac{\fB_{\Dir}\boxtimes S_{\cM_\cA\to \cM_\cA/\widehat\cA} \boxtimes S^{\text{LG}}_{H\subset G}}{G^{(0)}}\,.
\ee
Here, the field theory $S^{\text{LG}}_{H\subset G}$ is the LG theory spontaneously breaking the $G$ symmetry to $H$ symmetry. The field theory $S_{\cM_\cA\to \cM_\cA/\widehat\cA}$ is a phase transition from $\cM_{\cA}$ to $\cM_\cA/\widehat\cA$. We are not aware of a universal construction for $S_{\cM_\cA\to \cM_\cA/\widehat\cA}$, but we can rerun the same argument that we applied to $\cA \subset \cT$ to the pair $\wh{\cA}\subset \cM_\cA$. 
After rerunning this protocol several times, we may arrive at a theory where $\langle \widehat{\cA}\rangle= \cM_\cA$ i.e. is modular. In that case we propose to include scalars for each generating anyon of $\widehat{\cA}$. Here, a generating anyon  means a line that upon repeatedly fusing with itself generates all lines in $\widehat{\cA}$.
We will discuss them in examples concretely. 

In summary, the proliferation phase transition between $\cT\to \cT/\cA$ in terms of the SymTFT sandwich, is realized as follows 
\begin{equation}\label{eq:mainresult}
\boxed{\begin{split}
  &\cP_{\cT\to \cT/\cA}\\
   = &\SymTFTEXTRA{\frac{\fB_{\Dir}\boxtimes  S_{\cM_\cA\to \cM_\cA/\widehat\cA} \boxtimes S^{\text{LG}}_{H\subset G}}{G^{(0)}}}{\fZ(G)}{\fB^{\cN_\cA}_{\Neu (G)}}\,.
\end{split}}
\end{equation}
Note that when $\cM_\cA$ and $\widehat\cA$ are trivial, the transition reduces to the transition of type (i), as discussed in  section \ref{sec:typei}. On the other hand, when $H$ and $G$ are trivial, the transition reduces to type (iii). 

This concludes the general theory how proliferation transitions are realized in the SymTFT for a general condensable algebra $\cA$ in an MTC $\cT$. We now turn to illustrating this with examples, abelian and non-abelian.

\section{Proliferating Abelian Anyons}
\label{sec:Zousuke}

We first apply our general proposal to abelian anyons, with no anomaly -- the anomalous case will be discussed later. 
This case has been studied also in \cite{Cheng:2026qax}, and so serves as a point of comparison. 
For abelian TOs, the condensable algebras $\cA$ are finite abelian groups, and so $\cA \cong \Rep(A)$, with $A\cong \Z_{n_1} \times \cdots \times \Z_{n_s}$ decomposed into $s$ simple factors. 
Our  proposal allows for a systematic rewriting of $\cT$ in terms of a concrete continuum abelian Chern-Simons-matter theory, that describes the proliferation transition \eqref{TtoTA}. 
Note that for abelian anyons, all algebras are of type (i). 

Given $\cT$ and $\cA=\langle\cA\rangle= \Rep(A)$, we first need to compute the input data for the SymTFT. 
The SymTFT is the (3+1)d $A$-gauge theory. The symmetry boundary is $\Neu(a)$ and the physical boundary is stacked with 
\be
\cN= {\cT\over \Rep(A) }\,.
\ee
Note this is preciselt in the general discussion $\cN_\cA= (\Rep(A))'/\Rep(A)$.
% the topological order $\cN_\cA=\frac{\langle \cA \rangle{\Rep(A)} $ is the reduced TO $\cT/\cA$, while $\cM_\cA$ is trivial
We can then represent $\cT$ as
\be 
\cT=  
\SymTFTS{\Bsym_{\Neu(A)} }{ \fZ(A) }{ \fB ^{\cN_\cA}_{\Neu(A)} } \,.
\ee 
On the symmetry boundary we have dynamical $A$ gauge fields, that we  couple to dynamical scalar fields $\phi_a$ (one for each generator of $A$). The $A$-Wilson lines, can terminate non-topologically on $\phi_a$. In the massive phase we integrate out $\phi_a$, hence in the IR the lines are not endable, recovering $\fB_{\Neu(A)}$. In the Higgs phase $\langle\phi_a\rangle \not=0 $ condense, producing topological local operators in the IR, on which the Wilson lines can end. Hence the long distance limit of the boundary condition is $\fB_{\Dir(A)}$. Thus we have
 \begin{equation}
     \SymTFTS{\Bsym_{\Dir(A)} }{\fZ(A) }{\fB ^{\cN_\cA}_{\Neu(A)} }=\cN_\cA=\cT/\cA \,.
 \end{equation}
We can now write out a field theory derived from the SymTFT:
the (3+1)d $G$ gauge theory can be represented in terms of continuum $U(1)$ gauge fields as 
 \begin{equation}
     L=\sum_{i=1}^s \frac{in_s}{2\pi} b_i \, da_i \,.
 \end{equation}
 Here $b_i$ are 2-form $U(1)$ gauge fields, and $a_i$ 1-form $U(1)$ gauge fields for each simple factor of $A$. The symmetry boundary, before coupling with the scalars and its boundary Lagrangian are 
 \begin{equation}
 b_i\big|_{{\Bsym_{\Neu(A)}}}=dc_i\,, \qquad 
        L_{{\Bsym_{\Neu(A)}}}^{(\text{top})}=\sum_{i=1}^s \frac{in_i}{2\pi} c_i \, da_i \,,
 \end{equation}
 while $a_i$ are dynamical. Here $b_i$ is pure gauge, and $c_i$ is a 1-form $U(1)$ gauge field on the boundary. 
 We then introduces $s$-many complex scalar fields $\phi_i$ with Lagrangian
 \begin{equation}
     L_{\text{matter}}=\sum _{i=1}^s |D_{a_i}\phi _i|^2 + V\big( \left\{\phi_i\right\}\big) \ , \ \ \ \  D_{a_i}=\partial -ia_i \ .
 \end{equation}
Finally, on the physical boundary we can represent $\cN_\cA$ as an abelian Chern-Simons theory with K-matrix $K_\cA$, whose Lagrangian we denote by $L_{K_\cA}$. This has a natural coupling with $A$ gauge fields, hence the action arising from the physical boundary is 
\be 
L_{\Bphys}=L_{K_\cA}\left(\left\{a_i\right\}\right)\,.
\ee 
The dynamical theory arising from slab compactification has action
given by 
\be 
L_{\text{matter}}+
L_{\Bsym_{\Neu(A)}}^{(\text{top})} + L_{\Bphys}\,.
\ee
To be precise, one needs to always add appropriate monopole operators to this Lagrangian (see also \cite{Cheng:2026qax}). We will leave this implicit in the various examples.

\paragraf{Example: $\cT= U(1)_8$.}
Consider $\cT=U(1)_8$, and the algebra $\cA=\Rep(\Z_2)$. We have $\cN_\cA=U(1)_8/\Rep(\Z_2)\cong U(1)_2$, whose Lagrangian coupled with $a$ is $\frac{2i}{4\pi}x \, dx+\frac{i}{2\pi}a\, dx$. The final transition theory is 
\begin{equation}
    |D_a\phi|^2 +V(\phi) +\frac{2i}{2\pi}c\, da +\frac{2i}{4\pi}x \, dx +\frac{i}{2\pi}a \, dx \ .
\end{equation}
For $m_\phi^2>>0$ we remain the last 3 terms that define a $U(1)^3$ Chern-Simons theory equivalent to $U(1)_8$. 
For $m_\phi^2<<0$, $a$ is frozen, hence we remain with $\frac{2i}{4\pi}x\, dx$.

To see the dynamics of the theory, we need to add back the monopole operator $M_c$ for $c$ gauge field. The bare $M_c$ carries charge $2$ under $U(1)_a$, so it should be dressed by $\phi^{2\dagger}$ to make the monopole operator gauge invariant, i.e. $M_c \phi^{2\dagger}$. Integrating out $c$ makes $a$ into a $\Z_2$ gauge field, and the theory is a topological manipulation of a $\Z_2$ gauged XY model deformed by $\phi^2+h.c.$, i.e. the Ising$^*$ transition. In summary, the transition is second order, and in the Ising$^*$ universality class.

\paragraf{Example: $\cT=D(\Z_4)$.}
It is useful to consider an example where the abelian symmetry group has multiple simple factors, as this illustrates that even in the abelian case, the BC can carry some non-trivial SPT stacking. 
$\cT= D(\Z_4)$ has lines $e^{n_e}m^{n_m}$ with spin 
\be 
\theta_{n_e,n_m}=\exp{\left(\frac{2\pi in_en_m}{4}\right)}\,.
\ee 
Consider the condensable  algebra 
\be 
    \cA=1\oplus e^2 \oplus m^2 \oplus e^2m^2 \cong \Rep(\Z_2\times \Z_2) \, .
\ee 
The SymTFT is the $\Z_2 \times \Z_2$ gauge theory $\fZ(\Z_2\times \Z_2)$, explicitly 
\begin{equation}
    \frac{2i}{2\pi}a _1db_1 +\frac{2i}{2\pi}a_2db_2 \ .
\end{equation}
The symmetry boundary is $\Neu(\Z_2\times \Z_2)$, and  its coupling with two scalar fields $\phi_1,\phi _2$ are as in the general discussion. 

Now consider the physical boundary: Given that $\cT/\cA$ is trivial, one might conclude that the physical boundary is the bare Neumann. This is incorrect. In fact the enrichment with a 0-form symmetry  can be non-trivial even for a trivial TQFT: once enriched with $A=\Z_2\times \Z_2$, $\cT/\cA$ is a non-trivial $\Z_2\times \Z_2$ SPT given by a type II cocycle $\omega_{II} \in H^3(\Z_2,\times \Z_2,U(1))=\Z_2^3$:
\begin{equation}\label{eq:Z2xZ2 SPT}
    \frac{i}{2\pi}\int A_1 dA_2 \, .
\end{equation}
Here $A_1,A_2$ are $U(1)$ gauge fields with $\pi  \Z$ valued holonomies, that serve as background fields for the two $\Z_2$ factors. {A way to derive \eqref{eq:Z2xZ2 SPT} is the following. The $\Z_2\times \Z_2$ symmetry arises as the dual symmetry after gauging $\Rep(\Z_2\times \Z_2)$, hence the gauged theory coupled with background field is 
\begin{equation*}
     \frac{4i}{2\pi}xdy +\frac{2i}{2\pi}x c_x^{(2)} +\frac{2i}{2\pi}yc_y^{(2)}+\frac{i}{2\pi} c_x^{(2)}A_1 +\frac{i}{2\pi}c_y^{(2)}A_2 \ .
\end{equation*}
Here $c_x^{(2)}$ and $c_y^{(2)}$ are 2-form gauge fields whose holonomies are valued in $\pi \Z$. Integrating them out sets $2x=A_1, 2y=A_2$, and plugging this back we obtain \eqref{eq:Z2xZ2 SPT}}. In \cite{Bhardwaj:2024qiv} these were called $\Neu(A)^{\omega_{II}}$ modified Neumann BCs, which are obtained by stacking Dirichlet with an SPT and gauging $A$. 

We conclude that $\cN_\cA$ is given by this SPT, hence the physical boundary condition is the gauged SPT.
The SymTFT is then 
\be
\cT=  \SymTFTS{\Bsym_{\Neu(\Z_2\times\Z_2)}}{ \fZ (\Z_2\times\Z_2)}{\Neu(\Z_2 \times \Z_2)^{\omega_{II}}} \,.
\ee
We can write the complete field theory, including the Higgs fields, as follows:
\begin{equation}
\ba
     |D_{a_1}\phi_1|^2+|D_{a_2}\phi _2|^2+ V(\phi _1,\phi _2) + \\
     +\frac{2i}{2\pi}c_1da_1 +\frac{2i}{2\pi}c_2da_2+\frac{i}{2\pi}a_1 da_2 \ .
\ea    
\end{equation}
For $m^2_{\phi_1},m_{\phi_2}^2>>0$ we remain with a dual Chern-Simons description of $D(\Z_4)$. On the other hand in the complete Higgs phase the theory is trivial, as the symmetry boundary becomes $\Bsym_\Dir$ and the SymTFT compactifies to $\Vec$.

\section{Proliferating Non-Abelian Anyons: $\cT=D(S_3)$}
\label{sec:S3}

Group quantum doubles $D(G)$ for non-abelian finite groups $G$ are the simplest examples of non-abelian TOs. We will here focus on 
 $G=S_3= \Z_3^{a}\rtimes \Z_2^b$. However, a general analysis for any finite $G$ where all possible condensable algebras are known and the associated generated subcategories can be explicitly determined will appear in \cite{YGSSN}. 
 
For $D(S_3)$ the pure charges are irreps: the trivial $1$, the sign $1_-$, and the 2d irrep $E$. The fluxes are the conjugacy classes $[a]$ and $[b]$. The condensable algebras $\cA$ are listed in figure \ref{fig:S3Hasse}. The generated subcategories $\langle \cA \rangle$ are summarized in table \ref{tab:DS3}. We will discuss the proliferation transitions for all the condensable algebras of this TO.

The following algebras are {\bf type (i) subcategories $\Rep(G)$}: the non-maximal algebra 
$\Rep(\Z_2) = 1\oplus 1_-$, and the two Lagrangian algebras 
\be 
\Rep(S_3)^e = 1\oplus 1_-\oplus 2E\,,\qquad 
\Rep(S_3)^m = 1\oplus 1_-\oplus 2[a]\,.
\ee
For each of these we will  construct the SymTFT \eqref{TSymTFT}. The relevant scalars that we need to couple to are listed  in table \ref{tab:DS3}.

\paragraf{Type (i) Algebra: $\cA=1\oplus 1_-= \Rep(\Z_2)$.}
We apply the general results in  section \ref{sec:typei} to the current case.

The generated subcategory $\langle \cA\rangle$ in $D(S_3)$
is $\Rep(\Z_2)$, hence the algebra is of type (i).  The SymTFT is $\fZ(\Z_2)$--the 3+1d $\Z_2$ gauge theory.   The symmetry boundary is $\fB_{\Neu(\Z_2)}$ realizing the $\Rep(\Z_2)$ symmetry. The physical boundary is $\fB_{\Neu(\Z_2)}^{D(\Z_3)}$ since $\cN_\cA=\langle \cA\rangle'/\Rep(\Z_2)= D(\Z_3)$. Putting the above information together, we see the SymTFT sandwich for $D(S_3)$ is 
\begin{equation}
  D(S_3)=  \SymTFTS{\fB_{\Neu(\Z_2)}}{\fZ(\Z_2)}{\fB_{\Neu(\Z_2)}^{D(\Z_3)}} \,,
\end{equation}
and the SymTFT for the theory $\cT/\cA= D(\Z_3)/\Rep(\Z_2) = D(\Z_3)$ is
\begin{equation}
    D(\Z_3)= \SymTFTS{\fB_{\Dir(\Z_2)}}{\fZ(\Z_2)}{\fB_{\Neu(\Z_2)}^{D(\Z_3)}}\,.
\end{equation}
The proliferation transition \eqref{eq:LG gauged} becomes 
\begin{equation}\label{eq:transitionZ2}
    \cP_{D(S_3)\to D(\Z_2)} =  \SymTFTEXTRA{\frac{\fB_{\Dir}\boxtimes S^{\text{LG}}_{\Z_1\subset \Z_2}}{\Z_2^{(0)}}}{\fZ(\Z_2)}{\fB_{\Neu(\Z_2)}^{D(\Z_3)}}\,.
\end{equation}

%%%%%%%%%%%%%%%%%%%%%%%%%%%%%%%%%%%%%%
\definecolor{c1}{rgb}{0.0, 0.72, 0.92}
\definecolor{c2}{rgb}{1, 0.57, 0.13}
\definecolor{c3}{rgb}{0.8, 0.3, 1}

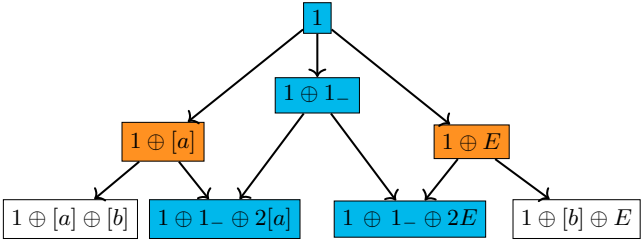
\begin{figure}
\centering
\adjustbox{max width=\columnwidth}{
\begin{tikzpicture}[vertex/.style={draw}]
\begin{scope}[shift={(0,0)}, scale=0.85]
\node[vertex, fill= c1] (1)  at (0,0.25) {${1}$};
\node[vertex, fill=c2] (2)  at (-2.5,-1.75) {$1\oplus [a] $};
\node[vertex, fill= c1] (3) at (0, -1) {$1 \oplus 1_-$} ;
\node[vertex, fill= c2] (4) at (2.5, -1.75) {$1 \oplus E$} ;
\node[vertex] (5) at (-4, -3) {$
1 \oplus [a]\oplus [b]$} ;
\node[vertex, fill= c1] (6) at (-1.5, -3) {$1 \oplus 1_- \oplus 2[a] $} ;
\node[vertex, fill= c1] (7) at (1.5, -3) {$1 \,\oplus\, 1_- \oplus 2E $};
\node[vertex] (8) at (4.2, -3) {$1 \oplus [b] \oplus E$} ;
\draw[thick, ->] (1) edge  node[label=left:] {} (2);
\draw[thick, ->] (1) edge  node[label=left:] {} (3);
\draw[thick, ->] (1) edge  node[label=left:] {} (4);
\draw[thick, ->] (2) edge  node[label=left:] {} (5);
\draw[thick, ->] (2) edge  node[label=left:] {} (6);
\draw[thick, ->](3) edge  node[label=left:] {} (6);
\draw[thick, ->] (3) edge  node[label=left:] {} (7);
\draw[thick, ->] (4) edge  node[label=left:] {} (7);
\draw[thick, ->] (4) edge  node[label=left:] {} (8);
\end{scope}
\end{tikzpicture}
}
\caption{The condensable algebras of $D(S_3)$, partially ordered by the subalgebra relation. The cyan ones form subcategories, orange ones have a generated sub-category $\Rep (G)$, the uncolored  are mixed algebras of type (iii). \label{fig:S3Hasse}}
\end{figure}

\begin{table}[t]
$$
\small
\begin{array}{|l|c|l|l|c|l|}\hline
\cA & \text{type} & \langle\cA\rangle & \phi & H & \cT/\cA \\\hline\hline
1\oplus 1_- & \text{(i)} &\cA= \Rep (\Z_2) & \phi_-  & 1 & D(\Z_3)\\
1\oplus E & \text{(i)} & \Rep (S_3)^e & \phi_E & \Z_2 & D(\Z_2)\\
1\oplus [a] & \text{(i)} & \Rep (S_3)^m & \phi_{[a]} & \Z_2' & D(\Z_2)\\
1\oplus 1_-\oplus 2E & \text{(i)} &  \cA= \Rep (S_3)^e & (\phi_E,\phi_-) & 1& \Vec\\
1\oplus 1_-\oplus 2[a] & \text{(i)} &\cA= \Rep (S_3)^m & (\phi_{[a]},\phi_-) & 1 & \Vec\\
1\oplus E\oplus [b] & \text{(iii)} & D(S_3) & \text{2-step} & - & \Vec\\
1\oplus [a]\oplus [b] & \text{(iii)}& D(S_3) & \text{2-step} & - & \Vec\\\hline
\end{array}
$$
\caption{The non-trivial condensable algebras $\cA$ of $D(S_3)$, their generated sub-category, order parameter, vacuum stabilizer $H$ and endpoint.  \label{tab:DS3}}
\end{table}

To be more concrete about the proliferation transition \eqref{eq:transitionZ2}, we first describe the theory $S^{\text{LG}}_{\Z_1\subset \Z_2}$. The theory has a real scalar $\phi_-$ in the sign representation of $\Z_2$, i.e. it transforms under $\Z_2$ as $\phi_- \to - \phi_-$. The Lagrangian is
\begin{equation}\label{eq:11-}
    L_{\Z_1\subset \Z_2}^{\text{LG}}=(\partial \phi_-)^2 + m^2 \phi_-^2 + \lambda \phi_-^4 \,.
\end{equation}
The effect of tensoring $\fB_{\Dir}$ and gauging the $\Z_2^{(0)}$ symmetry is to couple the LG theory \eqref{eq:11-} to a flat $\Z_2$ gauge field $a$, and making it dynamical. Then $\phi_-$ lives on the end point of the $1_-$ line of $\Rep(\Z_2)$. Feeding this into the SymTFT sandwich \eqref{eq:transitionZ2} amounts to performing a topological manipulation of the LG transition, 
\begin{equation}
    \frac{S^{\text{LG}}_{\Z_2\to \Z_1}\boxtimes D(\Z_3)}{\Z_2^{(0)}}\,,
\end{equation}
where in the denominator, we gauge the $\Z_2$ symmetry by summing over the  flat $\Z_2$ bundle that acts as  charge conjugation in $D(\Z_3)$, and acts on the scalar as $\phi_-\to -\phi_-$.

\paragraf{Type (i) Algebras: $\Rep(S_3)^e$ and $\Rep(S_3)^m$.} These two cases are essentially the same, and we only discuss the $e$ type. 
The generated subcategory $\langle \cA\rangle$ is  $\Rep (S_3)$, hence the algebra is of type (i). Its centralizer is $\langle \cA \rangle'=\Rep(S_3)$. The SymTFT is $\fZ(S_3)$--the 3+1d $S_3$ gauge theory. The symmetry boundary is $\fB_{\Neu(S_3)}$, realizing the $\Rep(S_3)$ symmetry.
The physical boundary is $\fB_{\Neu(S_3)}$ since $\cN_\cA= \langle \cA \rangle'/\Rep(S_3) = \Vec$. Putting the above information together, we find the SymTFT sandwich for $D(S_3)$ is 
\be \label{Ponyo}
  D(S_3)=  \SymTFTS{\fB_{\Neu(S_3)}}{\fZ(S_3)}{\fB_{\Neu(S_3)}}\,,
\ee 
and the SymTFT sandwich for the $\Rep(S_3)^e$ gauged theory $D(S_3)/\Rep(S_3)^e= \Vec$ is 
\begin{equation}
    \Vec = \SymTFTS{\fB_{\Dir(S_3)}}{\fZ(S_3)}{\fB_{\Neu(S_3)}}\,.
\end{equation}
The proliferation transition \eqref{eq:LG gauged} becomes 
\begin{equation}\label{eq:DS3Vec}
    \cP_{D(S_3) \stackrel{\Rep(S_3)^e}{\rightarrow} \Vec} = \SymTFTEXTRA{\frac{\fB_{\Dir}\boxtimes S_{\Z_1\subset S_3}^{\text{LG}}}{S_3^{(0)}}}{\fZ(S_3)}{\fB_{\Neu(S_3)}}\,.
\end{equation}

To be more concrete about the proliferation transition \eqref{eq:DS3Vec}, first describe the theory $S_{\Z_1\subset S_3}^{\text{LG}}$. 
The theory has two scalar multiplets $\phi_E\in \mathbb{R}^2, \phi_-\in \mathbb{R}$. 
It is useful to define a complex scalar $\Phi:= \phi_{E,1}+i\phi_{E,2}\in \mathbb{C}$. The $S_3$ action is 
\begin{equation}
    \Z_2^b: 
    \begin{cases}
        \phi_-\to -\phi_-\\
        \Phi\to  \Phi ^*
    \end{cases} \qquad 
    \Z_3^a: 
    \begin{cases}
        \phi_-\to \phi_-\\
        \Phi \to e^{\frac{2\pi i}{3}} \Phi \,.
    \end{cases}
\end{equation}
The Lagrangian of the LG theory for $S_3$ spontaneously breaking transition is \cite{Codello:2020mnt, Chatterjee:2022tyg} 
\begin{equation}\label{eq:S3LG}
\begin{split}
    L_{\Z_1\subset S_3}^{\text{LG}} = &(\partial \phi_-)^2 + |\partial \Phi|^2 + m_{_-} ^2\phi_-^2+\lambda _{_-} \phi_-^4\\
    &+m_{_E}^2|\Phi|^2+ \lambda _{_E} |\Phi |^4+ \gamma \phi ^2_- |\Phi|^2 \\&+ \alpha \text{Re}(\Phi ^3)+\beta \phi_-\text{Im}(\Phi ^3) \,.
\end{split}
\end{equation}
Indeed the potential is $S_3$ invariant, and in the Higgs phase $m_{_-}^2,m_{_E}^2<<0$, the generic vevs $\langle \phi_-\rangle, \langle \Phi \rangle $ Higgses $S_3$ completely. The phase transition is generically first order due to the cubic term.

The effect of tensoring $\fB_{\Dir}$ and gauging the $S_3^{(0)}$ symmetry is to couple the LG theory \eqref{eq:S3LG} to a flat $S_3$ gauge bundle and summing it over. Then $\phi_-$ lives on the end point of the $1_-$ line of $\Rep(S_3)$, and $\Phi$ lives on the end point of $E$ line. Feeding this to the SymTFT sandwich \eqref{eq:DS3Vec} amounts to performing a topological manipulation of the LG transition, 
\begin{equation}
    \frac{S^{\text{LG}}_{\Z_1\subset S_3}}{S_3^{(0)}}\,.
\end{equation}
This is also known as the LG$^*$ transition.

We also remark that this transition can be intuitively understood as two sequential transitions by the transitivity property \eqref{residualdimension}, 
\begin{equation}\label{eq:sequential1}
    1\oplus 1_- \oplus 2 E: \quad D(S_3)\ \xrightarrow{\ 1\oplus E\ }\ D(\Z_2)\ \xrightarrow{\ 1\oplus e\ }\ \Vec
\end{equation}
this should be contrasted with a similar transition which realizes type (iii) algebra in \eqref{eq:sequential2}.

\bigskip

There are two {\bf type (i) with non-trivial $H$} algebras: 
 $1\oplus E$ and $1\oplus [a]$ do not close under fusion. They generate $\Rep (S_3)^e$ and $\Rep (S_3)^m$, respectively, and are  instances of type (i) with $H$ non-trivial. Again these work very similarly and we consider only the  the $e$ one. 

\paragraf{Type (i) Algebra $\cA=1\oplus E$ with $H=\Z_2$.}
For $\cA=1\oplus E=\Fun(S_3/\Z_2)$ and 
$\langle \cA \rangle = \Rep(S_3)^e$. 
Thus $H=\Z_2$. Its centralizer is $\langle \cA \rangle'=\Rep(S_3)$. The SymTFT is again a 3+1d $S_3$ gauge theory $\fZ(S_3)$, and the symmetry boundary and physical boundaries are both  $\fB_{\Neu(S_3)}$. Hence the SymTFT sandwich for $D(S_3)$ is the same as \eqref{Ponyo}. The SymTFT sandwich for the gauged theory $D(S_3)/(1\oplus E) = D(\Z_2)$ is 
\be
\label{pigpig}
D(\Z_2)=\SymTFTS{\fB_{\Neu(\Z_2)}}{\ \fZ(S_3)\ }{\fB_{\Neu(S_3)}}\,.
\ee
The proliferation transition \eqref{eq:LG gauged} becomes 
\begin{equation}\label{eq:DS3DZ2}
    \cP_{D(S_3) \stackrel{1\oplus E}{\longrightarrow} D(\Z_2)} = \SymTFTEXTRA{\frac{\fB_{\Dir}\boxtimes S_{\Z_2\subset S_3}^{\text{LG}}}{S_3^{(0)}}}{\fZ(S_3)}{\fB_{\Neu(S_3)}}\,.
\end{equation}

The proliferation transition \eqref{eq:DS3DZ2} can be described more concretely with  the theory $S_{\Z_2\subset S_3}^{\text{LG}}$. It contains a single doublet $\phi_E$ in the 2d irrep, which can again be packaged to a complex scalar $\Phi$. The Lagrangian is 
\begin{equation}
    L_{\Z_2\subset S_3}^{\text{LG}} = |\partial \Phi|^2 + m^2|\Phi|^2+ \lambda |\Phi |^4+ \alpha \text{Re}(\Phi ^3) \,.
\end{equation}
Note that this is related to \eqref{eq:S3LG} by deleting the $\phi_-$. The potential is automatically $S_3$ invariant. To see the property of the Higgs phase, we assume $\alpha <0$ without loss of generality. While the generic $\Phi$ has trivial stabilizer, in the broken phase  ($m^2< m_c^2=\alpha^2/4\lambda$) the potential has three minima
\be
\langle \Phi \rangle _k =v_* e^{\frac{2\pi ik}{3}} \ , \ \ \ \ k=0,1,2 % 
\ee
so that
\be
\text{Stab}\left(\langle \Phi \rangle _k \right)=\langle a^k b a^{-k}\rangle \cong \Z_2 \ .
\ee 
So the vacua are labeled by the coset $S_3/\Z_2$, and are rotated by $\Z_3$. The model is the Landau-Ginzburg description of the ferromagnetic 3-state Potts model \cite{Zia:1975ha, Amit:1976pz}. Our model is the $S_3$ gauging of that, 
\begin{equation}
    \frac{S^{\text{LG}}_{\Z_2\subset S_3 }}{S_3^{(0)}}\,.
\end{equation}

In absence of fine tuning of the cubic term (i.e. $\alpha =0$, for which there would be an accidental $O(2)$ symmetry), the transition is first order \cite{DJAmit_1974}.

\bigskip
The remaining two algebras are $1\oplus [b] \oplus E$ and $1\oplus [a] \oplus [b]$. They are related by $\Z_2$ automorphism, and hence work similarly. We will focus on $1\oplus [b] \oplus E$. 

\paragraf{Type (iii) Algebra: $\cA=1\oplus [b] \oplus E$. } The generated subcategory $\langle \cA\rangle$ is $D(S_3)$, which has trivial Muger center. As a consequence, the SymTFT is just $2\Vec$, and the physical boundary is also $\Vec$. This shows that the algebra is of type (iii).  Hence the SymTFT sandwich construction is not helpful for constructing the proliferation theory. 

However, by the transitivity property \eqref{residualdimension}, we observe that gauging this type (iii) algebra can be realized sequential gauging, 
\begin{equation}\label{eq:sequential2}
    1\oplus E\oplus[b]:\quad D(S_3)\ \xrightarrow{\ 1\oplus E\ }\ D(\Z_2)\ \xrightarrow{\ 1\oplus m\ }\ \Vec\,.
\end{equation}
This is similar to \eqref{eq:sequential1}.
Since each step can be realized by proliferation transition, we can concatinate the two transitions to get a one-step transition. This perspective will be elaborated further in \cite{YGSSN}.

\paragraf{Comments about group quantum doubles $\cT=D(G)$.}
In \cite{YGSSN} the general non-abelian $G$ case will be discussed. Here we will make a few comments: there is always the electric Lagrangian algebra, which corresponds to all charges ending, which is the Dirichlet BC for $D(G)$. This is always the $\Rep (G)$ type (i) algebras . Normal subgroups $H$ of this will also provide $\Rep (G/N)$ type (i) algebras. 
In general there will also be 
 mixed algebras we should note that the step-wise procedure  works more generally: 
for $G$ solvable, every condensable algebra of $D(G)$ admits a complete filtration by such steps (whereas for $G$ simple and non-abelian the magnetic Lagrangian algebra $\bigoplus_{[g]}([g],1)$ admits no intermediate stage).
The first step is always obtained by  considering the restriction to $\Fun(G/H)$, which is a subalgebra, whose generated category is (a sub-category) of the charges $\Rep (G)$. Condensing this gives rise to the first step: $D(G)\to D(H)$. One can keep iterating this process, but in the intermedate steps, the reduced TOs might have twists, i.e. $D(K)^\omega$.  The general analysis for group quantum double proliferations will appear in \cite{YGSSN}.

\section{Proliferating Non-Abelian Anyons: $\cT=SU(2)_k$ and its Double}
\label{sec:SU2k}

The Chern-Simons theories $SU(2)_k$ and their doubles
$\cT_k=SU(2)_k\boxtimes SU(2)_{-k}$ realize all three types of condensable
algebras (i)-(iii), and is thus a very suitable class of examples. 
Furthermore, which type of algebra occurs depends on $k$ in a very simple way: 
The anyons of
$SU(2)_k$ are the {isospins} $(j)$, $j=0, \half,1, \cdots, {k\over 2}$, with quantum dimensions and spins
\be
\label{su2kdata}
d_j=\frac{\sin\big((2j+1)\pi/(k+2)\big)}{\sin\big(\pi/(k+2)\big)}\,,\qquad
\theta_j=e^{2\pi i\, j(j+1)/(k+2)}\,.
\ee
There is a unique non-tirival invertible line $J=({k\over 2})$. Its spin
\be
\label{Jtwist}
\theta_J=e^{i\pi k/2}\,, 
\ee
makes $J$ a boson for $4\mid k$, a fermion for $k\equiv 2\bmod 4$ and a semion for $k$ odd. 
The entire discussion below is therefore uniform
in $k$ modulo $4$. The exceptional levels which have additional structures are  $k=10,16,28$. 

\subsection{The Chiral Theory $\cT=SU(2)_k$}
\label{sec:SU2kchiral}

The condensable algebras of $SU(2)_k$ follow the ADE classification of modular invariants \cite{Kirillov:2001ost}
\be
\label{ADElistmain}
\ba
\cA_A&=(0)\,, &&\text{every }k\,,\cr
\cA_D&=(0)\oplus\big(\tfrac k2\big)\,, &&4\mid k\,,\cr
\cA_{E_6}&=(0)\oplus(3)\,, &&k=10\,,\cr
\cA_{E_8}&=(0)\oplus(5)\oplus(9)\oplus(14)\,,\quad &&k=28\,.
\ea
\ee

\paragraf{Proliferation Transitions for $4\mid k$.}
The $\cA_D$ algebra 
is of type (i): for $4\mid k$ the current $J$ is a boson
braiding trivially with itself, and thus 
\be
\langle\cA_D\rangle=\cE_{\cA_D}=\langle J\rangle\cong\Rep (\Z_2)\,.
\ee
The SymTFT is the (3+1)d $\Z_2$ gauge theory, the symmetry boundary is $\Neu(\Z_2)$, and the physical boundary is non-minimal, stacked with 
\be \label{QkDef}
SO(3)_{k/2}=\frac{SU(2)_k}{(0)\oplus\big(\tfrac k2\big)} \,.
\ee
This is precisely the gauged theory.
This allows us to rewrite $SU(2)_k$ as 
\be
\label{ADsandwich}
SU(2)_k=\SymTFTS{\fB_{\Neu (\Z_2)}}{\ \fZ(\Z_2)\ }{\fB^{SO(3)_{k/2}}_{\Neu (\Z_2)}} 
\,.
\ee
and also the gauged theory $SO(3)_{k/2}$ as 
\be
\label{ADendpoint}
SO(3)_{k/2}=\SymTFTS{\fB_{\Dir}}{\ \fZ(\Z_2)\ }{\fB^{SO(3)_{k/2}}_{\Neu (\Z_2)}}\,.
\ee
The transition is 
\be \ba 
   & \cP_{SU(2)_k\to SO(3)_{k/2}}\cr 
   &=\SymTFTEXTRA{\frac{\fB_{\Dir} \boxtimes S_{\Z_1\subset \Z_2}^{\text{LG}}}{\Z_2^{(0)}}}{\ \fZ(\Z_2)\ }{\fB^{SO(3)_{k/2}}_{\Neu (\Z_2)}}\,.
\ea
\ee
The LG theory $S_{\Z_1\subset \Z_2}^{\text{LG}}$ is the same as Ising transition. 

Shrinking the sandwich, this means that the phase transition is a topological manipulation of Ising transition 
\begin{equation}
    \frac{S^{\text{LG}}_{\Z_1\subset \Z_2}\boxtimes SO(3)_{k/2}}{\Z_2^{(0)}}\,.
\end{equation}

\paragraf{Exceptional cases: $k=10,  28$.}
The generated subcategory for the two exceptional algebras for $k=10,28$ are
\begin{equation}
    \langle\cA_{E_6}\rangle=\PSU(2)_{10}, \quad \langle\cA_{E_8}\rangle=\PSU(2)_{28}\,,
\end{equation}
since $(3)\otimes(3)$ at $k=10$ and $(5)\otimes(5)$ at $k=28$ contain $(1)$,
which generates every integer isospin.
Note that $\PSU(2)_{k}$ is different from $SO(3)_{k/2}$ -- the former is obtained by deleting the half isospin anyons from $SU(2)_k$, while the latter is obtained by gauging the algebra $\cA_D$ from $SU(2)_k$.

Both subcategories are pre-modular--they have non-trivial M\"uger center. For
$k=10$ it is generated by a {\bf transparent fermion} $(5)$ and is
super-Tannakian, so this requires the stacking construction for anomalous symmetries in section \ref{sec:anomalousgeneral}. We will not discuss this case in the current section.

For $k=28$ the M\"uger center is
$\Rep (\Z_2)$, generated by the boson $(14)$. Hence the SymTFT is $\fZ(\Z_2)$. The symmetry boundary is $\fB_{\Neu(\Z_2)}^{SO(3)_{14}}$.  The physical boundary is $\fB_{\Neu(\Z_2)}$. Combining the ingredients together, we find the SymTFT sandwich for $SU(2)_{28}$ is
\begin{equation}
    SU(2)_{28} = \SymTFTS{\fB_{\Neu(\Z_2)}^{SO(3)_{14}}}{\fZ(\Z_2)}{\fB_{\Neu(\Z_2)}}\,.
\end{equation}

To write down the $\cA_{E_8}$ gauged theory $SU(2)_{28}/\cA_{E_8} = (G_2)_1$ as SymTFT sandwich, we first note that $\cA_{E_8}\cap \Rep(\Z_2) = \Rep(\Z_2) = \Fun(\Z_2)$, hence $H=1$ in \eqref{transparentpart}. Substituting into \eqref{eq:T/Ageneral}, we find the SymTFT sandwich for $(G_2)_1$ is 
\begin{equation}
    (G_2)_1 = \SymTFTS{\fB_{\Dir}\boxtimes (G_2)_1}{ \fZ(\Z_2)}{\fB_{\Neu(\Z_2)}}\,.
\end{equation}
Note that the symmetry boundary can also be rewritten as $(\fB_{\Dir}\boxtimes (G_2)_1\boxtimes Q_{\Z_2})/\Z_2$. 
Then according to the general proposal for the transition theory \eqref{eq:mainresult}, we need to find a Higgs transition 
\be \label{E8residual}
    SO(3)_{14}\to (G_2)_1\,.
\ee
We can identify this transition as an inverse Higgsing transition as follows. 

\paragraf{Exceptional Transitions from Conformal Embeddings.}
The exceptional algebras have a uniform origin: the level-1 conformal
embeddings of $SU(2)$ \cite{Schellekens:1986mb, Bais:1986zs}
\be\ba 
SU(2)_4 &\subset SU(3)_1\cr 
SU(2)_{10}&\subset \Spin(5)_1\cr 
SU(2)_{28}&\subset (G_2)_1 \,,
\ea\ee 
defined by the equality of Sugawara stress-tensors of the current algebras $T_{G_1}=T_{SU(2)_k}$. 
The
condensable algebra is the vacuum branching rule of the embedding: 
this
reproduces $\cA_D$ at $k=4$ and $\cA_{E_6}$, $\cA_{E_8}$ at $k=10,28$. In all three cases the embedded $SU(2)$ is the
principal one: 
i.e. the $\mathfrak{sl}(2)$ subalgebra generated by a regular nilpotent element, unique up to conjugation, under which $\mathfrak{g}$ decomposes into spins given by the exponents $m_i$ of $\mathfrak g$ \cite{MR114875}. E.g. for 
$G_2$ it is the $SO(3)$ under which the $\mathbf 7$ is the spin-3 representation.
Denoting by $V_j$ the spin-$j$ representation and by
$T(V_j)=\tfrac13 j(j+1)(2j+1)$ its Dynkin index,
\be
\begin{array}{c|c|c|c|c}
k & G_1 & \text{adj}(\mathfrak g)|_{\rm pr} & k=x_{\rm emb}=\sum_i {T(V_{m_i})\over h^\vee} & \Sigma \\\hline
4 & SU(3)_1 & V_1\oplus V_2 & 12/3 & \mathbf{27}\\
10 & \Spin(5)_1 & V_1\oplus V_3 & 30/3 & \mathbf{35}\\
28 & (G_2)_1 & V_1\oplus V_5 & 112/4 & \mathbf{77}
\end{array}
\ee
with $m_i$ the exponents of $\mathfrak g$, $x_{\rm emb}=k$, and $\Sigma$ the
representation $V_{2\theta}\subset\text{Sym}^2(\text{adj})$, $\theta$ the
highest root. The transition theory is $G_1$ Chern-Simons coupled to a scalar
$\Sigma\in V_{2\theta}$. The vev is canonical: the Casimir tensor
$\Sigma_{\rm pr}=(\sum_a t_a\otimes t_a)_{\rm traceless}$ of the embedded
subalgebra, which is the unique principal singlet in $V_{2\theta}$ and has
stabilizer exactly the principal $SU(2)$; since the invariant subspace is
one-dimensional, the principal stratum is an extremum of every
$G_1$-invariant potential. For $k=4,28$ the exponents are integer spins, the
embedding factors through $SO(3)$, and the Higgsed phase is the quotient
$SO(3)_{k/2}$ of \eqref{QkDef}: at $k=28$ this is the residual transition
$SO(3)_{14}\to (G_2)_1$ of \eqref{E8residual}. 

For $k=10$ the spinor
$\mathbf 4$ restricts to $V_{3/2}$, the center survives, and the Higgs frame
realizes the fermionic transition $SU(2)_{10}\to\Spin(5)_1$ in a single step,
consistent with the super-Tannakian M\"uger center and the absence of a
condensable sub-algebra. Finally, since
$\text{Sym}^2(\mathfrak g)^{SU(2)}$ is spanned by the Killing form
and $\Sigma_{\rm pr}$, the defining identity $T_{G_1}=T_{SU(2)_k}$ is
equivalent to
$: \Sigma_{\rm pr}^{ab}\,J_a J_b :\ \propto\ T_{G_1}$, 
i.e. the order-parameter bilinear of the currents is the stress tensor
itself: the transition has no independent local order parameter, as expected for the proliferation of anyons.

\subsection{The Double $\cT_k=SU(2)_k\boxtimes SU(2)_{-k}$}
\label{sec:SU2kdouble}

The MTCs $\cT_k=SU(2)_k\boxtimes SU(2)_{-k}$ are non-chiral, and the anyons are pairs of spins $(j_1,j_2)$. Depending on $k$, there can be different condensable algebras and proliferation transitions. 
The algebras and their Hasse diagrams for generic (non-exceptional) $k$ are listed in figure \ref{fig:SU2kHasse}.
We discuss only the algebras that do not factorize into chiral algebras that we discussed in section \ref{sec:SU2k}.

\subsubsection{Type (i) Algebra $\Rep (\Z_2)$} 

One condensable algebra exists for all values of $k$.
Let us define 
\be
\label{abcmain}
a=\big(\tfrac k2,0\big)\,,\qquad b=\big(0,\tfrac k2\big)\,,\qquad c=\big(\tfrac k2,\tfrac k2\big)\,,
\ee
with $\theta_a=\bar\theta_b=\theta_J$ and $\theta_c=1$: in particular the
diagonal algebra 
\be 
\cA_J=(0,0)\oplus\left({k\over 2},{k\over 2}\right)\cong\Rep (\Z_2)
\ee 
is condensable for every $k$ and it is a type (i) algebra.
Yet again, the SymTFT is the (3+1)d $\Z_2$ gauge theory with physical boundary stacked with 
\be 
\cN_k={\cT_k\over \cA_J}=SO(4)_{k,-k} \,.
\ee 
The SymTFT sandwich of $\cT_k$ is 
\be
\label{AJsandwich}
\cT_k= \SymTFTS{\fB_{\Neu (\Z_2)}}{\ \fZ(\Z_2)\ }{\fB^{\cN_k}_{\Neu (\Z_2)}}\,.
\ee
The SymTFT sandwich of the gauged theory $\cT_k/\cA$ is 
\begin{equation}
   { \cT_k/\cA}= \SymTFTS{\fB_{\Dir}}{\ \fZ(\Z_2)\ }{\fB^{\cN_k}_{\Neu (\Z_2)}}\,.
\end{equation}
The phase transition is 
\begin{equation}
    \cP_{\cT_k\to \cN_k} = 
    \SymTFTS{\frac{\fB_{\Dir}\boxtimes S^{\text{LG}}_{\Z_1\subset \Z_2}}{\Z_2^{(0)}}}{\ \fZ(\Z_2)\ }{\fB^{\cN_k}_{\Neu (\Z_2)}}\,.
\end{equation}

Concretely, there is a  single $\Z_2$-odd scalar $\phi_c$ driving the Higgsing
$\Neu (\Z_2)\to\Dir$ on the boundary. In terms of the $SO(4)_{k,-k}$ the matter is just a real scalar coupled to the $\Z_2$ gauge field, and the transition is in the Ising$^*$ universality class.

%%%%%%%%%%%%%%%%%%%%%%%%%%%%%%%%%%%%%%%%%%%%%%%%%%%%%%%%%%%%%%%%%%%%%%%%
\begin{figure*}
\centering
{
\adjustbox{max width=\textwidth}{
\begin{tikzpicture}[vertex/.style={draw, text=black}, scale=1.0]
%%% panel 1: k odd
\begin{scope}[shift={(0,0)}]
\node at (0,1.1) {$k$ odd, $k\geq 3$};
\node[vertex, fill=c1] (o1) at (0,0) {$1$};
\node[vertex, fill=c1] (oJ) at (-1.3,-1.4) {$\cA_J$};
\node[vertex] (oA) at (1.3,-1.4) {$\cA_{\ad}$};
\node[vertex, fill=c2] (oL) at (0,-2.8) {$\cL_{\diag}$};
\draw[thick, ->] (o1) edge (oJ);
\draw[thick, ->] (o1) edge (oA);
\draw[thick, ->] (oJ) edge (oL);
\draw[thick, ->] (oA) edge (oL);
\end{scope}
%%% panel 2: k = 2 mod 4
\begin{scope}[shift={(5.0,0)}]
\node at (0,1.1) {$k\equiv 2 \bmod 4$, $k\geq 6$};
\node[vertex, fill=c1] (t1) at (0,0) {$1$};
\node[vertex, fill=c1] (tJ) at (0,-1.4) {$\cA_J$};
\node[vertex, fill=c2] (tA) at (0,-2.8) {$\cA_{\ad}$};
\node[vertex, fill=c2] (tL) at (-1.3,-4.2) {$\cL_{\diag}$};
\node[vertex, fill=c2] (tS) at (1.3,-4.2) {$\cL_{\sigma}$};
\draw[thick, ->] (t1) edge (tJ);
\draw[thick, ->] (tJ) edge (tA);
\draw[thick, ->] (tA) edge (tL);
\draw[thick, ->] (tA) edge (tS);
\end{scope}
%%% panel 3: 4 | k
\begin{scope}[shift={(11.0,0)}]
\node at (0,1.1) {$4\mid k$};
\node[vertex, fill=c1] (f1) at (0,0) {$1$};
\node[vertex, fill=c1] (fa) at (-2.8,-1.4) {$1\oplus a$};
\node[vertex, fill=c1] (fb) at (0,-1.4) {$1\oplus b$};
\node[vertex, fill=c1] (fc) at (2.8,-1.4) {$\cA_J$};
\node[vertex, fill=c1] (fK) at (-1.4,-2.8) {$\cK$};
\node[vertex, fill=c2] (fA) at (2.8,-2.8) {$\cA_{\ad}$};
\node[vertex, fill=c2] (fD) at (-1.4,-4.2) {$\cL_{D}$};
\node[vertex, fill=c2] (fL) at (2.8,-4.2) {$\cL_{\diag}$};
\draw[thick, ->] (f1) edge (fa);
\draw[thick, ->] (f1) edge (fc);
\draw[thick, ->] (f1) edge (fb);
\draw[thick, ->] (fa) edge (fK);
\draw[thick, ->] (fc) edge (fK);
\draw[thick, ->] (fb) edge (fK);
\draw[thick, ->] (fc) edge (fA);
\draw[thick, ->] (fK) edge (fD);
\draw[thick, ->] (fA) edge (fD);
\draw[thick, ->] (fA) edge (fL);
\end{scope}
\end{tikzpicture}
}}
\caption{The Hasse diagrams of condensable algebras of $\cT_k=SU(2)_k\boxtimes SU(2)_{-k}$
for generic $k$ odd, $k=2 \mod  4$ and $k=0\mod 4$, excluding the exceptional cases $k=10,16,28$. 
cyan are type (i), orange type (ii), uncolored type (iii) algebras, respectively. 
The algebras are: $\cA_J=1\oplus c$, 
$\cA_{\ad}=\oplus_{j\in\Z}(j,j)$, the two chiral
$D$-type algebras $1\oplus a$ and $1\oplus b$, 
$\cK=1\oplus a\oplus b\oplus c$. Finally there are the Lagrangian algebras: 
$\cL_\diag=\oplus_j(j,j)$, 
$\cA_D=\cA_{\ad}\otimes(1\oplus a)$, and 
$\cL_\sigma$ of \eqref{sigmaLagrangianmain}. 
At $k\equiv 2\bmod 4$, $\cA_{\ad}$ is
super-Tannakian, so its type (ii) only for 
$\langle 1 \oplus c\rangle$. \label{fig:SU2kHasse}}
\end{figure*}
%%%%%%%%%%%%%%%%%%%%%%%%%%%%%%%%%%%%%%%%%%%%%%%%%%%%%%%%%%%%%%%%%%%%%%%%

\subsubsection{Type (ii) Diagonal Lagrangian Algebra $\cL_{\diag}$}

The diagonal (all spins, integer and half-integer) Lagrangian algebra  is
$\cL_\diag = \bigoplus_{j} (j,j)$. 
The generated subcategory $\langle \cL_{\text{diag}}\rangle$ contains all the lines $(j_1, j_2)$ with $j_1=j_2$ mod $2$. Its M\"uger center is $\Rep(\Z_2)$, hence the SymTFT is $\fZ(\Z_2)$. The symmetry boundary is $\fB_{\Neu(\Z_2)}^{SO(4)_{k,-k}}$, where $SO(4)_{k, -k}= \langle \cL_{\text{diag}}\rangle / \Rep(\Z_2)$. The physical boundary is $\fB_{\Neu(\Z_2)}$. The SymTFT sandwich is 
\begin{equation}
    SU(2)_k \boxtimes SU(2)_{-k} = \SymTFTS{\fB_{\Neu(\Z_2)}^{SO(4)_{k,-k}}}{\fZ(\Z_2)}{\fB_{\Neu(\Z_2)}}
\end{equation}
After gauging $\cL_{\text{diag}}$, we get a trivially gapped theory. The symmetry boundary becomes 
\begin{equation}
    \fB_{\Dir} = \frac{\fB_{\Dir} \boxtimes Q_{\Z_2} }{\Z_2^{(0)}} \,.
\end{equation}
The SymTFT sandwich for the transition theory is
\be 
\ba 
&    \cP_{\cT_k\to \cT_k/\cL_{\text{diag}}} = \cr 
&\SymTFTS{\frac{\fB_{\Dir(\Z_2)}\boxtimes S_{SO(4)_{k,-k}\to \Vec} \boxtimes S^{\text{LG}}_{\Z_1\subset \Z_2}}{\Z_2^{(0)}}}{\fZ(\Z_2)}{\fB_{\Neu(\Z_2)}}\,.
\ea
\ee
The field theory $S_{SO(4)_{k,-k}\to \Vec}$ is a transition from $SO(4)_{k,-k}$ to a trivially gapped theory. It is accompanied by the LG transition $\Z_2$ preserving to $\Z_2$ SSB phase.

While it is difficult to find the transition $S_{SO(4)_{k,-k}\to \Vec}$, it is easy to find the two transition combined together, i.e. $S_{SO(4)_{k,-k}\to \Z_2\text{SSB}}$. This is achieved by turning on a scalar in the vector representation of $SO(4)$, which was studied in detail in \cite{Ji:2026yfj}.

It is useful to derive this transition from our general framework, but in this case, we should also note that there is a direct way to get to it, as done in  \cite{Ji:2026yfj}.  
The transition theory is obtained by coupling a scalar $\phi$ in the bifundamental representation of $SU(2)\times SU(2)$ to the gauge fields and condensing the scalar: 
\be\label{eq:singletrans}
L_{SU(2)_k \boxtimes SU(2)_{-k}}+|D \phi |^2 +m^2 |\phi|^2 + \lambda |\phi|^4 \,.
\ee
When $m^2\gg 0$, the theory flows to $SU(2)_k \boxtimes SU(2)_{-k}$. When $m^2\ll 0$, $SU(2)\times SU(2)$ gauge group is Higgsed down to the diagonal subgroup whose level is 0. Therefore the Higgs phase is trivially gapped phase.

\subsubsection{Type (ii) and (iii) Integer-spin Diagonal Algebra $\cA_{\ad}$}
\label{sec:integerspindiagonal}

The integer-spin diagonal algebra and its generated subcategory is 
\be
\label{Aadmain}
\cA_{\ad}=\bigoplus_{j\in\Z}\,(j,j)\,,\qquad \langle\cA_{\ad}\rangle = \PSU(2)_k\boxtimes\PSU(2)_{-k} \,,
\ee
which is a subalgebra of $\cL_\diag = \oplus_{\text{all }j} (j,j)$. 
For even $k$ we have $\cE_{\cA_{\ad}}=\{1,a,b,c\}$, while for odd $k$ it is trivial, $\cE_{\cA_{\ad}}=\Vec$ and we have the following algebra types:
\be
\label{Aadtrichotomy}
\ba
k\ \text{odd}: &\ \ \text{(iii)}\,, &\ \cT_k/\cA_{\ad}&=D(\Z_2^\omega)\cr
k\equiv0\bmod4: &\ \ \text{(ii)}\,, &\ \cT_k/\cA_{\ad}&=D(\Z_2)\cr
k\equiv2\bmod4: &\ \ \text{(ii) (fermionic)}\,, &\ \cT_k/\cA_{\ad}&=D(\Z_2)\,,
\ea
\ee
with $D(\Z_2^\omega)$ the double semion. The $k\equiv 2\bmod4$ case includes transparent fermions in the M\"uger center and we will not discuss this here. 
We now derive the proliferation transitions for the $k$ odd and multiple of $4$ case, $\cA=\cA_{\ad}$ for this discussion:

\paragraf{Case $k$ odd.} For $k$ odd  the algebra is type  (iii) i.e. modular and the theory factorizes as 
\be 
\cT_k=\big(\PSU(2)_k\boxtimes\PSU(2)_{-k}\big)\boxtimes D(\Z_2^\omega)\,.
\ee

The SymTFT sandwich for $\cT_k$ is 
\begin{equation}
    \cT_k = \SymTFTS{\fB^{\PSU(2)_k \boxtimes \PSU(2)_{-k}}}{\Vec}{\fB^{D(\Z_2^\omega)}} \,.
\end{equation}
The proliferation transition drives the symmetry boundary to a trivial boundary, 
\begin{equation}
    \cP_{\cT_k \to \cT_k/\cA_{\ad}} = \SymTFTS{S_{\PSU(2)_k \boxtimes \PSU(2)_{-k}\to \Vec}}{\Vec}{\fB^{D(\Z_2^\omega)}}\,.
\end{equation}
The boundary phase transition is realized by coupling a scalar $\phi_{(1,1)}$ in the 3 dimensional representation of $SU(2)$ for each factor, i.e. $(j,j)=(1,1)$, and condense it.

\paragraf{Case $4\mid k$.} 
For $k\equiv0\bmod4$ we have the generated subcategory  
\be
\langle\cA\rangle = \PSU(2)_k\boxtimes\PSU(2)_{-k}
\ee
with M\"uger center 
\be
\cE_\cA= \Rep(\Z_2\times \Z_2) \,.
\ee
The SymTFT is the 3+1d $\Z_2\times \Z_2$ gauge theory and we have a non-minimal symmetry boundary with (see (\ref{QkDef}))
\be
\cM_\cA = {\langle\cA\rangle \over \cE_\cA} = SO(3)_{k/2} \boxtimes SO(3)_{-k/2} \,.
\ee
The SymTFT rewriting of $\cT_k$ induced by this condensable algebra is 
\be\label{TikQQ} 
\cT_k=  \SymTFTS{\fB_{\Neu(\Z_2^2)}^{SO(3)_{k/2} \boxtimes SO(3)_{-k/2} }}{\fZ(\Z_2^2)}{\fB_{\Neu(\Z_2^2)}} \,.
\ee

Next, we determine the SymTFT sandwich data after transition -- see the discussion in section \ref{sec:typeiii}: the transparent anyons in $\cA$ are
\be
\label{Aadtransparentpart}
\ba
\cA \cap\cE_{\cA}&= (0,0)\oplus\left({k\over 2},{k\over 2}\right)\cr
&=\Fun\big(G/H\big) \cong \Rep (\Z_2)\,,
\ea
\ee
with $H=\Z_2$ the diagonal subgroup. Then in the second step we need to identify the image of $\cA$  in $\cM_\cA$ under the map $F: \langle \cA \rangle \to \cM_\cA$. This generates a diagonal algebra $\widehat{\cA}$ in $\cM_\cA$. 
Its dimension makes it Lagrangian. To see this, note first that the integer spins are precisely the lines that braid trivially with $J=({k\over 2})$, i.e. $\PSU(2)_k=\langle J\rangle'$, and thus theorem \ref{thm:Muger-gluing} gives $\dim \PSU(2)_k=\half\dim SU(2)_k$. Since $\dim\cM_\cA=\dim\langle\cA\rangle/|G|$ and $|G|=4$,
\be
\label{Aaddims}
\ba
\dim\cM_\cA&={\big(\dim \PSU(2)_k\big)^2\over 4}
={1\over 16}\big(\dim SU(2)_k\big)^2\cr
&=\big(\dim SO(3)_{k/2}\big)^2\,,
\ea
\ee
where we used $\dim SO(3)_{k/2}={1\over 4}\dim SU(2)_k$, which follows from \eqref{quotientdimension} since the algebra $(0)\oplus\big({k\over 2}\big)$ of \eqref{QkDef} has dimension 2. This is of course just $\cM_\cA= SO(3)_{k/2} \times SO(3)_{-k/2}$. The algebra itself has dimension
\be
\label{Aaddim}
\dim\cA_{\ad}=\sum_{j\in\Z}d_j^2=\dim \PSU(2)_k=2\dim SO(3)_{k/2}\,,
\ee
since the anyon $(j,j)$ has quantum dimension $d_j^2$, and $[G:H]=2$ by \eqref{Aadtransparentpart}. Therefore
\be
\label{Aadhatdim}
\dim\wh\cA={\dim\cA\over [G:H]}=\dim SO(3)_{k/2}=\sqrt{\dim\cM_\cA}\,,
\ee
which is the Lagrangian condition in $\cM_\cA$, and we have
\be
\cM_\cA/\widehat{\cA} = \Vec\,.
\ee
The above results imply the symmetry boundary in \eqref{TikQQ} is replaced by
\begin{equation}
\frac{\fB_\Dir \boxtimes \Vec\boxtimes Q_{\Z_2^2/\Z_2}}{\Z_2^{(0)}\times \Z_2^{(0)}} \,.
\end{equation}
Feeding it into the SymTFT sandwich gives $D(\Z_2)$ as expected.

The phase transition is given in terms of the SymTFT sandwich as
\be 
\ba 
&    \cP_{\cT_k\to \cT_k/\cA_{\ad}} = \cr 
&\SymTFTEXTRA{{\fB_{\Dir}\boxtimes S_{ \cM_\cA \to \Vec}  \boxtimes S^{\text{LG}}_{\Z_2\subset\Z_2^2}
\over \Z_2^{(0)}}
}{\fZ(\Z_2^2)}{\fB_{\Neu(\Z_2^2)}}\,.
\ea
\ee
The phase transition $S_{ \cM_\cA \to \Vec}$ is realized by coupling a scalar in the bi-vector, i.e. $(1,1)$, representation of $SO(3)\times SO(3)$, and condensing it.  

In the special case $k=4$, since $\cT_4/\Z_2= D(S_3)$, the transition is related to those in section \ref{sec:S3}.

\subsubsection{Type (ii) Lagrangian Algebra $\cL_D$} 
For $k=4m$ we have the D-type Lagrangian algebra 
\be 
\ba 
 \cL_D=&2(m, m) \cr 
& \oplus \bigoplus_{j=0}^{m-1}[(j, j) \oplus(2 m-j, 2 m-j)  \cr 
 &\qquad \qquad  \oplus(j, 2 m-j) \oplus(2 m-j, j)  ]\,. 
\ea
\ee

To find the correct field theory, we use our SymTFT formalism. The generated subcategory of $\cL_{D}$ is
\be
\langle\cL_D \rangle = \PSU(2)_k \times \PSU(2)_{-k}
\ee
so that the M\"uger center is 
\be 
\cE_{\cA} = \Rep (\Z_2 \times \Z_2)\,,
\ee
with SymTFT the 3+1d $\Z_2\times \Z_2$ gauge theory. Since 
\be
\cM_\cA = {\langle\cL_D\rangle \over \cE_\cA} = SO(3)_{2m}\boxtimes SO(3)_{-2m}\,,
\ee
the non-minimal symmetry boundary is $\fB_{\Neu(\Z_2\times \Z_2)}^{\cM_\cA}$. The physical boundary is $\fB_{\Neu(\Z_2\times \Z_2)}$. It implies that the SymTFT sandwich is 
\be\label{TikQQ} 
\cT_k=  \SymTFTS{\fB_{\Neu(\Z_2^2)}^{\cM_\cA}}{\fZ(\Z_2^2)}{\Neu(\Z_2^2)} \,.
\ee

% i.e. we can use the same presentation of $\cT_k$ as in (\ref{TikQQ}). 
However, different to $\cA_{\ad}$ however, algebra $\cA_\D$ contains all of the transparent bosons
\be
\cA_D\cap \cE_\cA= \cE_\cA \,.
\ee
And so the first step in the transition is a full Higgsing of $\Neu(\Z_2\times\Z_2)$ to $\Dir (\Z_2\times \Z_2)$, and $H=\Z_1$. 

The second step is as follows:  we need to determine the image $\widehat{\cA}$ of $\cA$ under $F:\langle\cA_D \rangle \to \cM_\cA$. 
Again we can argue using dimensions that the resulting algebra $\widehat{\cA}$ is Lagrangian and 
$\cM_\cA/\widehat{\cA} = \Vec$:
Again the dimensions force $\wh\cA$ to be Lagrangian. The generated subcategory, and with it $\cM_\cA$, is the same as for $\cA_{\ad}$, so $\dim\cM_\cA=(\dim SO(3)_{2m})^2$ as in \eqref{Aaddims}. The algebra itself has dimension
\be
\label{ADdim}
\ba
\dim\cA_D&=2d_m^2+4\sum_{j=0}^{m-1}d_j^2=2\dim \PSU(2)_k\cr
&=\dim SU(2)_k=4\dim SO(3)_{2m}\,,
\ea
\ee
where we used $d_{2m-j}=d_j$, and $\big(\dim\cA_D\big)^2=\dim\cT_k$ confirms that it is Lagrangian in $\cT_k$. Since $H=1$, so that $[G:H]=|G|=4$ and
\be
\label{ADhatdim}
\dim\wh\cA={\dim\cA_D\over [G:H]}=\dim SO(3)_{2m}=\sqrt{\dim\cM_\cA}\,.
\ee
Correspondingly $F(\cA_D)$ splits into four connected components, one for each of the four vacua of the completely Higgsed $\Z_2\times\Z_2$, and the vev selects the one in the identity coset. So $\wh\cA$ is Lagrangian in $\cM_\cA$ and $\cM_\cA/\wh\cA=\Vec$.
The  SymTFT sandwich for gauged theory is 
\be
\Vec=\SymTFTEXTRA{\frac{\fB_{\Dir}\boxtimes \Vec\boxtimes Q_{\Z_2^2}}{\Z_2^{(0)}\times \Z_2^{(0)}}}{\fZ(\Z_2^2)}{\fB_{\Neu(\Z_2^2)}}\,.
\ee
The whole proliferation transition is described in terms of the SymTFT 
\be 
\ba 
&    \cP_{\cT_k\to \cT_k/\cL_{D}} = \cr 
&\SymTFTEXTRA{{\fB_{\Dir(\Z_2^2)}\boxtimes S_{\cM_\cA\to \Vec} \boxtimes S^{\text{LG}}_{\Z_1\subset \Z_2^2}
\over (\Z_2 \times \Z_2)^{(0)}}
}{\fZ(\Z_2^2)}{\fB_{\Neu(\Z_2^2)}}\,.
\ea
\ee
Again the phase transition $S_{\cM_\cA\to \Vec}$ is realized by coupling a scalar in the bi-vector, i.e. $(1,1)$, representation of $SO(3)\times SO(3)$, and condense it.

\subsubsection{Discussion}

\paragraf{Transparent fermions.}
For $k\equiv 2\mod 4$, consider $\cA_{\ad}$. The anyons $a$ and $b$ are  fermions, which becomes part of the M\"uger center  $\cE_{\cA}=\Rep (\Z_2\times\Z_2,z)$ of $\langle\cA_{\ad}\rangle$ ("super-Tannakian M\"uger center"). We generally did not discuss super-Tannakian cases. However, here 
we can nevertheless  study the proliferation transition,  because we can consider the smaller subcategory $\langle c\rangle\cong\Rep (\Z_2)$, which has a bosonic M\"uger center 
$\cA_J\subset\cA_{\ad}$ is a subalgebra: the slab \eqref{AJsandwich} supplies
the first step and the residual algebra is condensed in $\cZ(\PSU(2)_k)$.  
Another case with transparent fermions is $\cA_{E_6}$ at $k=10$, that however has no condensable subalgebra. 

\paragraf{Transitions with different multiplicative structures.} 
A second feature is specific to $k\equiv 2\mod 4$: the map
$\sigma$ fixing the integer spins and sending $j\mapsto\tfrac k2-j$ on the
half-integer ones preserves fusion, $S$ and $T$, and its graph
\be
\label{sigmaLagrangianmain}
\cL_\sigma=\bigoplus_j\big(j,\sigma(j)\big)
\ee
is a second Lagrangian algebra \cite{Cappelli:1986hf}. $\cL_\sigma$ has the same generated subcategory as $\cL_\diag$ and the same endpoint of the proliferation transition as
$\cL_\diag$. Nevertheless these phases should differ in terms of the  multiplicative structure of anyons, and it would be interesting to explore this further, in conjunction with twin algebra \cite{Warman:2026gfz, Gai:2026hjk} proliferation transitions.

\section{Proliferating Anomalous Anyons}
\label{sec:Ano}

So far we focused on condensable algebras $\cA$, which by definition are non-anomalous. An anomalous 1-form symmetry of $\cT$ cannot be gauged within $\cT$ by itself. It can nevertheless be combined or stacked with an auxiliary TQFT $\cX$ carrying the opposite anomaly, after which an anomaly-free diagonal can be gauged. The choice of $\cX$ is arbitrary, apart from its anomaly-properties. On the dynamical QFT part of the story, one can couple with a scalar matter sector, which upon Higgsing, produces a phase in which the anomalous lines appear to be gauged. Neither the auxiliary TQFT nor the dynamical realization of this operation is unique.  Here we extend the SymTFT construction of the previous sections using a canonical choice selected by the generated sub-category.  The resulting operation is a {\bf generalized gauging or generalized proliferation} rather than an anyon condensation intrinsic to $\cT$.

\subsection{SymTFT Construction for Anomalous Symmetries}
\label{sec:Factorization}

Let $X\in\cT$ be an arbitrary, not necessarily simple, object and define
\be
\label{anomalousframe}
\cB_X:=\langle X\rangle\subseteq\cT\,,\qquad
\cE_X:=\cZ_2(\cB_X)\,.
\ee
Here we again mean $\langle X \rangle$ to be the smallest sub-fusion category of $\cT$ that contains $X$. $\cB_{X}'$ is the centralizer of $\cB_X$ in $\cT$. 
However, now, we do not require $X$ to be an algebra. As before we restrict (for simplicity) to the case of the M\"uger center of this to be Tannakian, $\cE_X\cong\Rep(G)$, i.e. without transparent fermions.  
With this data, we now define the quotients
\be
\label{anomalousquotients}
\cT':=\frac{\cT}{\cE_X}\,,\qquad
\cM_X:=\frac{\cB_X}{\cE_X}\,,\qquad
\cN_X:=\frac{\cB_X'}{\cE_X}\,.
\ee
Both $\cM_X$ and $\cN_X$ are modular by construction, and since their dimensions are 
\be
\ba
\dim\cM_X&=\frac{\dim\cB_X}{|G|}\,,\cr
\dim\cN_X&=\frac{\dim\cT}{|G|\dim\cB_X}\,,\cr
\dim\cT'&=\frac{\dim\cT}{|G|^2}
=\dim\cM_X\,\dim\cN_X\,,
\ea
\ee
we can apply Theorem \ref{thm:Muger-gluing} to get a factorization of the MTC $\cT'$
\be
\label{anomalousfactorization}
\cT'\cong\cM_X\boxtimes\cN_X\,.
\ee
The SymTFT construction now lets us rewrite $\cT$ as follows: 
\be
\cT= \SymTFTS{\fB^{\cM_X}_{\Neu(G)}}{\fZ(G)}{\fB^{\cN_X}_{\Neu(G) }} \ .
\ee
We can then use this to construct a dynamical theory $S_{\cT+\Phi}$, via coupling scalar fields $\Phi=(\phi,\varphi)$ on the symmetry boundary. Here $\phi$ couples with the $G$ gauge field, while $\varphi$ denotes a set of scalar that couples with the MTC $\cM_X$ and is able, in the Higgs phase, to trivialize it. 
This last step requires a Chern-Simons representation of $\cM_X$ and needs to be worked out case by case. 

Our claim is that $S_{\cT+\Phi}$ is a field theory for the transition
\begin{equation}
    \cT\rightarrow \frac{\cT \boxtimes \overline{\cM_X}}{\cA^{\text{diag}}_X} \,,
\end{equation}
where $\cA^{\text{diag}}_X$ is a diagonal algebra constructed out of $\cB_X$ and its image through the projection functor $F: \cB_X \rightarrow \cM_X=\cB_X/\cE_X$. More precisely 
\begin{equation}
\label{anomalousgraphalgebra}
    \cA^{\text{diag}}_X
    =\bigoplus_{Y\in\Irr(\cB_X)}Y\boxtimes F(Y)^\vee\,,
\end{equation}
which has dimension $\dim\cA_X^{\text{diag}}=\dim\cB_X$ and contains
$\Rep(G)\boxtimes1$.  Condensing this subalgebra first gives
$\cM_X\boxtimes\cN_X\boxtimes\overline{\cM_X}$, and the residual algebra is
the ordinary diagonal Lagrangian algebra of
$\cM_X\boxtimes\overline{\cM_X}$.  Its condensation leaves precisely
$\cN_X$, i.e. we fine the endpoint SymTFT representation to be 
\be
\SymTFTS{\fB_\Dir}{\fZ(G)}{\fB_{\Neu(G)}^{\cN_{X}}} = \frac{\cT \boxtimes \overline{\cM_X}}{\cA^{\text{diag}}_X} = \cN_X \,.
\ee

\paragraf{Factorization.} This analysis provides a universal construction in the following sense: for any anyon $X$, it gives a prescription of a minimal TQFT $\overline{\cM_X}$, 
that we need to stack in order to then gauge $X$. We will argue that 
any other TQFT $\cX$, which we could stack and has the property that it cancels the anomaly, can be reduced, after a suitable condensation of an algebra $\widetilde{\cA}$, to factored into 
\be\label{donkeydoodle}
\cX/\widetilde{\cA}\cong \overline{\cM_X} \boxtimes \cX '\,,
\ee
where  $\cX'$ is an MTC that does not talk to $X$. 
This generalizes the minimal anomalous TQFT for abelian anyons in \cite{Hsin:2018vcg} to general non-abelian anyons. In fact, the precise meaning  of $\cX$ cancelling the anomaly, is that it must admit a braided ribbon functor $F_\cX : \cB_X \rightarrow \overline{\cX}$ 
such that 
\begin{equation}
    \bigoplus _{Y\in \cB_X} Y\boxtimes F(Y)^\vee 
\end{equation}
is a condensable algebra of $\cT \boxtimes \cX$. Concretely 
$F_\cX$ assigns to every line $Y\in \cB_\cX$ an auxiliary line with opposite braiding and twist, compatibly with fusion. The condensability of the algebra above is precisely the condition that the diagonal combination can be gauged.   Therefore $\text{Ker}\left( F\right) \subset \cE_X$, hence it induces an embedding
\be 
    \widehat{F} :\qquad  \cB_X/\cE_X \rightarrow \overline{\cX}/F(\cE_X) \,.
\ee
The algebra  $\overline{F(\cE_X)}$ appearing on the RHS is precisely what was denoted by $\widetilde{\cA}$ in (\ref{donkeydoodle}). 
Given that $\cM_X=\cB_X/\cE_X$ is modular, M\"uger factorization Theorem \ref{thm:Muger-gluing} gives $\cX/\overline{F(\cE_X)}=\overline{\cM_X} \boxtimes \cX'$, which is (\ref{donkeydoodle}).

We now demonstrate the general story above in concrete examples, both abelian and non-abelian, starting with the former.

\subsection{Anomalous Abelian Symmetries}

{For abelian TO $\cT$, in general $X=a_1\oplus \dots \oplus a_s$ where $a_i$ are simple abelian anyons, hence $\cB_X=\langle a_1,...,a_s\rangle$ is the subcategory whose fusion rules are governed by the finite abelian  group formed by the $a_i$. Without loss of generalities we can take the $a_i$ to be linearly independent, and if $n_i$ is the order of $a_i$, this group is $A=\Z_{n_1} \times \dots \times \Z_{n_s}$. The pre-modular structure of $\cB_X$ is determined by the spin $\theta : A \rightarrow U(1)$ induced from $\cT$, that in turns fixes 
\be
S_{\alpha, \beta}=\frac{\theta _{\alpha+\beta}}{\theta _\alpha \theta_\beta} \ .
\ee
If this bilinear form is non-degenerate, then $\cB_X$ is by itself modular and the TQFT is factorized $\cT=\cB_X \boxtimes \cN_X$.  Its M\"uger center $\cE_X$ is the abelian group:
\be
\cE_X=\left\{\alpha \in A \ | \theta_\alpha =1 \ ,  \ S_{\alpha,\beta}=1 \  \forall \beta \in A \right\} \ .
\ee
In turns $\cM_X =A/\cE_X$ inherits a modular structure from $\widehat{\theta} : A/\cE_X \rightarrow U(1)$, 
and defines a consistent TQFT by itself.}

For a single anyon $s=1$, with $X=a$ of order $n$, the discussion becomes as follows: the spin is
\begin{equation}
    \theta(a)=e^{\frac{2\pi ip}{2n}} \ , \ \ \ p\in \Z_{2n} \  , \ \ \ pn \in 2\Z \,.
\end{equation}
The last condition follows from $a^n=1$. $p\neq 0 \text{mod}(2n)$ represents the anomaly. This generated sub-category $\cB_a=\langle a\rangle $ is pre-modular with $\Z_n$ fusion rules and 
\be
\label{abelianquadraticform}
\theta_{a^r}=\exp\!\left(\frac{2\pi i p r^2}{2n}\right),
\qquad
S_{a^r,a^s}=\exp\!\left(\frac{2\pi i p rs}{n}\right).
\ee
We define 
\begin{equation}
    l:=\gcd(n,p) \ , \ \ \ \ n=ln' \ ,  \ \ \  \ p=lp' \ . 
\end{equation}
$\cB_a$ is modular only if $l=1$, otherwise it has M\"uger center 
$\cE_a =\langle a^{n'}\rangle \cong \Rep(\Z_l)$. We assume $p'n'\in 2\Z$ for the M\"uger center to be Tannakian (rather than super Tannakian). $\cM_a=\cB_a/\cE_a$ is an MTC with $n'$ invertible lines and the generator has spin $e^{\frac{2\pi i p'}{2n'}}$. This is the minimal theory defined in \cite{Hsin:2018vcg}
    \begin{equation}
        \cM_a=\cA^{n',p'} \ .
    \end{equation}
Notice that here we always get this minimal theory, also when $\text{gcd}(n,p)\neq 1$, by first gauging the non-anomalous symmetry (see \cite{Cheng:2025ube} for a related discussion).

The SymTFT provides the rewriting 
\begin{equation}
    \cT=\left(\fB_{\Neu(\Z_l)}^{\cA^{n',p'}} \ \big| \ \fZ(\Z_l) \ \big| \ \fB_{\Neu(\Z_l)}^{\cN_a} \right) \ .
\end{equation}
To mimic the condensation of $a$ we introduce, on the symmetry boundary, a scalar field $\phi$ that couples with the $\Z_l$ gauge field, as well as a set of scalar fields $(\varphi_\alpha)$ that are coupled only with $\cA^{n',p'}$ and trivialize completely that theory in the Higgs phase. In fact any minimal theory can be represented as an abelian Chern-Simons theory with a given K-matrix, so it is enough to introduce one scalar of unit charge under each $U(1)$ factor of the gauge group.

The complete Higgs phase correponds to a phase transition on the symmetry boundary
\begin{equation}
    \fB_{\Neu(\Z_l)}^{\cA^{n',p'}} \rightarrow
     \fB_{\Dir(\Z_l)}
\end{equation}
and therefore the field theory constructed in this way describes the transition
\begin{equation}
    \cT \xrightarrow{\langle \phi \rangle , \langle\varphi _\alpha \rangle} \cN_a=\frac{\cT' \boxtimes \overline{ \cA^{n',p'}}}{\Z_{n'}}= \frac{\cT \boxtimes \overline{ \cA^{n',p'}}}{\Z_{n}} \ .
\end{equation}
The last quotient is generated by the line $(a,y)$, with $y$ the generator of $\overline{ \cA^{n',p'}}$.

Notice that, except for the case $p=\pm 1$, this is a different transition from the one studied in \cite{Cheng:2026qax}, where the  choice $\cX= U(1)_{-pn}$ is made. That choice makes the coupling to a single scalar more natural, while in our framework the natural choice, which is minimal, is
\begin{equation}
    \cX= \overline{\cA^{n',p'}} \,,
\end{equation}
which however has the drawback that the transition theory will generically have multiple scalar fields.

\paragraf{Example: $\cT = U(1)_6$.} Consider $a = W_2$, so $n = 3$ and $p = 2$, hence $\ell = 1$ and the theory factorizes
\be
\label{U16factorization}
\ba
U(1)_6 &\,\cong\, \cM_a\boxtimes \cN_a= \cA^{3,2}\boxtimes U(1)_{-2}\,,\cr
\cA^{3,2} &= \{1, W_2, W_4\} \cong SU(3)_1 \,,\cr
{\cN_a} &= \{1, W_3\}\cong U(1)_{-2} \,.
\ea
\ee
%The Higgs phase is ${\cN_a} = U(1)_{-2}$. 
This decomposition favors a Lagrangian realization of $\cT$ as the $U(1)^3$ theory with
\be
\label{U16Kmatrix}
K = \begin{pmatrix} 2 & -1 & 0 \\ -1 & 2 & 0 \\ 0 & 0 & -2\end{pmatrix} \,,
\ee
the $A_2$ block realizing $SU(3)_1 = \cA^{3,2}$. We then couple this to two scalars $\varphi_{1,2}$ of charges $(1,0,0)$ and $(0,1,0)$: their condensation Higgses precisely the anomalous factor, realizing the transition
\begin{equation}
    U(1)_6 \xrightarrow{\langle \varphi_1 \rangle , \langle\varphi _2 \rangle} \frac{U(1)_6 \boxtimes \overline{SU(3)_1}}{\Z_3}=U(1)_{-2}
\end{equation}

Let us comment about the relation with \cite{Cheng:2026qax}. The question there is, in a sense, reversed relative to ours: starting from a given coupling with scalars, what is the Higgs phase? In \cite{Cheng:2026qax} the scalar coupling is somewhat minimal, using a single scalar, whose charge and Lagrangian presentation of $\cT$ selects an integer lift of $p$. In this specific example, coupling a single scalar of charge $2$ under $U(1)_6$ gives the Higgs phase
\be
\label{U16MengNati}
\frac{U(1)_6\boxtimes U(1)_{-6}}{\Z_3} =D(\Z_2^\omega)=U(1)_2\boxtimes U(1)_{-2}  \,,
\ee

 With respect to our Higgs phase, this has an additional decoupled $U(1)_2$ factor. Both are consistent proliferations of the same anyon $W_2$: they correspond to different UV completions, distinguished by the auxiliary sector. In our case we used the minimal TQFT, in \cite{Cheng:2026qax} the $U(1)_{-6}$ theory.

\paragraf{Example: $\cT = U(1)_{8}$.}
A richer example is $\cT=U(1)_{8}$ and $a=W_2$.  Then 
\be
\ba
n&=4\,,\qquad p=2\,,\qquad \ell=2\,,\cr
n'&=2\,,\qquad p'=1\,.
\ea
\ee
The M\"uger center and the three factors in \eqref{anomalousquotients} are
\be
\label{U1eightfactorization}
\ba
\cE_a&=\langle W_4\rangle\cong\Rep(\Z_2)\,,\cr
\cT'&=\frac{U(1)_8}{\Rep(\Z_2)}\cong U(1)_2\,,\cr
\cM_a&\cong U(1)_2\,,\qquad \cN_a\cong\Vec\,.
\ea
\ee
The SymTFT sandwich presentation is 
\begin{equation}
   \cT = \SymTFTS{\fB ^{U(1)_2}_{\Neu(\Z_2)}}{\fZ(\Z_2)}{\fB_{\Neu(\Z_2)}} \,.
\end{equation}
We couple the symmetry boundary with two scalars: $\phi$ couples with the $\Z_2$ gauge fields, while $\varphi$ with the $U(1)_2$ sector. This gives concrete field theory realization with two Higgs fields after the slab compactification
\be
\label{U1eighttransitionlagrangian}
\ba
L={}& |D_a\phi|^2+|D_x\varphi|^2+V(\phi,\varphi)\cr
&+\frac{2i}{2\pi}c\,da
+\frac{2i}{4\pi}x\,dx
+\frac{i}{2\pi}a\,dx\,.
\ea
\ee
The $K$-matrix in the basis $(a,c,x)$ is
\be
K=\begin{pmatrix}0&2&1\\2&0&0\\1&0&2\end{pmatrix}\,,
\ee
and it is easy to check that this is equivalent to $U(1)_8$, with the identifications
\be 
    W_1 =e^{i\int c} \ ,\quad W_2=e^{i\int x} \ , \quad  W_4=e^{i\int a} \,.
\ee
 The four phases are:
\be
\ba
\langle\phi\rangle=\langle\varphi\rangle=0 &:\quad  U(1)_8\,,\cr
\langle\phi\rangle\neq0,\ \langle\varphi\rangle=0 &: \quad U(1)_2\,,\cr
\langle\phi\rangle=0,\ \langle\varphi\rangle\neq0 &: \quad D(\Z_2)\,,\cr
\langle\phi\rangle\neq0,\ \langle\varphi\rangle\neq0 &: \quad \Vec\,.
\ea
\ee
The first one-field Higgsing is the non-anomalous $W_4$ transition of section \ref{sec:Zousuke}.  The second Higgses $x$ and leaves the level-two BF theory $(2i/2\pi)c\,d a $, namely the toric code. The complete Higgs phase is trivial, correctly reproducing
\begin{equation}
    \frac{U(1)_8 \boxtimes  U(1)_{-2}}{\Z_4}=\Vec \ .
\end{equation}

Let us consider two examples with multiple abelian anyons. 

 \paragraf{Example: $\cT=D(\Z_4)$ with multiple anyons.} 
In this example all anyons non-Anomalous by themselves, but carry a mixed anomaly. We consider $\cT=D(\Z_4)$ and $X=e\oplus m^2$. While $e$ and $m^2$ are both bosons, they have mutual non-trivial braiding hence they cannot be simultaneously gauged.

We have $\cB_X=\langle e,m^2\rangle\cong \Z_4 \times \Z_2$ with
\be
 S_{e^{x_1}m^{2y_1}, e^{x_2}m^{2y_2} }=(-1)^{x_1y_2+x_2y_1}\,.
\ee
Hence 
\be
\cE_X=\langle e^2\rangle \cong \Z_2 \ , \ \ \cM_X=D(\Z_2) \ , \ \ \cN_X=\Vec \ .
\ee
Notice that $\cM_X$ here is a higher rank generalization of the minimal TQFT of \cite{Hsin:2018vcg}, but differently form the single anyon case it can have anomaly free subgroups (see appendix of \cite{Antinucci:2022vyk}). The canonical generalized gauging procedure is:
\be
\cT \rightarrow \frac{\cT \boxtimes \overline{D(\Z_2)}}{\langle (e,\overline{e}) , (m^2,\overline{m}) \rangle}\cong \Vec  \ .
\ee
The SymTFT framework realizes $\cT=D(\Z_4)$ as:
\be
\SymTFTS{\fB_{\Neu(\Z_2)}^{D(\Z_2)}}{\fZ(\Z_2)}{\fB_{\Neu(\Z_2)}}=D(\Z_4) \ .
\ee
This favors the following Lagrangian presentation of $D(\Z_4)$ in terms of $U(1)$ gauge fields:
\be
\frac{2i}{2\pi}c da +\frac{2i}{2\pi} x dy +\frac{i}{2\pi}a dx \ .
\ee
The identification with the $D(\Z_4)$ anyons is
\be
e=e^{i\int x} \ , \ \  m=e^{i\int c} \ , \ \ e^2=e^{i\int a} \ , \ \  m^2=e^{i\int y} \ .
\ee

The dynamical theory is then obtained by coupling the symmetry boundary with three complex scalars $\phi, \varphi_{1,2}$. The first couples with the $\Z_2$ gauge field, while $\varphi _{1,2}$ couple with the $\cM_X=D(\Z_2)$ sector:
\be
\ba 
L={}& |D_a\phi|^2+|D_x\varphi_x|^2 +|D_y\varphi _2|^2+V(\phi,\varphi_1,\varphi_2)\cr
&+\frac{2i}{2\pi}c\,da
+\frac{2i}{2\pi}x\,dy
+\frac{i}{2\pi}a\,dx\,.
\ea
\ee
This field theory realizes a phase transition $D(\Z_4)\rightarrow \Vec$ by proliferating $e$ and $m^2$.

 \paragraf{Example: $\cT= U(1)^2$ CS with anomaly.}
 Consider the $U(1)^2$ theory with K-matrix 
\begin{equation}
K=\left(\begin{array}{cc}
8     & 4 \\
    4 & 8
\end{array}
\right) \ .
\end{equation}
We label the anyons by $(x,y)\sim (x+8,y+4)\sim (x+4,y+8)$. They generate the 1-form symmetry 
\be 
\langle (1,-1) , (1,0)\rangle \cong \Z_4 \times \Z_{12}\,.
\ee 
The spin is
\be
\theta_{x,y}=e^{\frac{2\pi i}{12}(x^2+y^2-xy)} \,.
\ee
We consider the anyon $X=(2,-2) \oplus (3,0)$, so that
\be
\cB_X=\left\{ (2r+3l,-2r) \ | \ r\in \Z_2 \ , l\in \Z_4 \right\}\cong \Z_2 \times \Z_4 \ .
\ee
The braiding on $\cB_X$ is 
\be
S_{(2r_1+3l_1,-2r_1) (2r_2+3l_2,-2r_2)}=e^{2\pi i \left(\frac{2}{3}r_1r_2+\frac{3}{2}l_1l_2\right)}
\ee
hence its M\"uger center is 
\be
\cE_X=\cZ_2(\cB_X) =\langle (6,0) \rangle \cong \Z_2\,.
\ee
After modding it out from $\cB_X$ there are 4 lines
\be
\ba
1\equiv (0,0)  \ :  \ \ \theta =1 \ \ \ \  &  \ \ \ \  s\equiv(-5,2) \ : \ \ \theta =i \\ \overline{s}\equiv (3,0) \ : \ \ \theta =-i \ \ \ \  &  \ \ \ \  s\overline{s}\equiv(-2,2) \ : \ \ \theta =1 \,,
\ea
\ee
and the resulting theory is the double semion
\be
\cM_X=\cB_X/\cE_X =D(\Z_2^\omega)=U(1)_2 \boxtimes U(1)_{-2}\,.
\ee
We also have $\cB_X'=\langle (2,0)\rangle \cong \Z_6$ and given that $(2,0)$ has spin $e^{2\frac{2\pi i}{3}}$ we have
\be
\cN_X=\cB_X'/\cE_X \cong SU(3)_1 \,.
\ee
The canonical generalized gauging procedure is 
\be
\cT \rightarrow \frac{ 
\cT \boxtimes \overline{D(\Z_2^\omega)}
}{\langle \left((1,-1);\overline{s}\right) , \left((3,0);s\right) \rangle 
}\cong SU(3)_1\,.
\ee
The SymTFT framework realizes $\cT$ as:
\be
\cT= \SymTFTS{\fB_{\Neu(\Z_2)}^{D(\Z_2^\omega)}}{\fZ(\Z_2)}{\fB_{\Neu(\Z_2)}^{SU(3)_1}} \,.
\ee 
The dynamical theory is then obtained by coupling the symmetry boundary with three complex scalars $\phi, \varphi_{1,2}$. The first couples with the $\Z_2$ gauge field, while $\varphi _{1,2}$ couple with the $\cM_X=D(\Z_2^\omega)$ sector. Finally, by realizing $SU(3)_1$ with K-matrix $\begin{pmatrix}
   2  &  1 \\
    1 & 2
\end{pmatrix}$, we can write its coupling with the $\Z_2$ gauge field explicitly as follows
\be 
\ba 
L={}& |D_a\phi|^2+|D_x\varphi_x|^2 +|D_y\varphi _2|^2+V(\phi,\varphi_1,\varphi_2)\cr
&+\frac{2i}{2\pi}c\,da
+\frac{2i}{2\pi}b\,dz
+\frac{i}{2\pi}b\,db+\frac{i}{2\pi}a\,db \cr
& +\frac{i}{2\pi} \alpha\,  da+\frac{i}{2\pi}\alpha \, d\alpha +\frac{i}{2\pi}\beta \, d\beta +\frac{i}{2\pi}\alpha\, d\beta \ .
\ea
\ee
This realizes the phase transition $\cT\rightarrow SU(3)_1$. In fact in the massive phase we remain with only the topological terms, and the theory can be shown to be equivalent to $\cT$, while in the complete Higgs phase (after adding the appropriate monopoles) only the $\alpha,\beta$ sector survives, and that is $SU(3)_1$.

\subsection{Anomalous Non-Abelian Anyon Proliferation}
\label{sec:anomalousgeneral}

The general proposal in section \ref{sec:Factorization} applies to any anyon in an MTC $\cT$, including non-abelian topological orders. 

Technically, however a concrete, field theoretic realization of the Higgsing requires a presentation of the TQFT $\cM_X$ to admit a Chern-Simons gauge presentation $\text{CS}[H,k]$, together with representations $R_\lambda$ containing the relevant Wilson lines such that generic scalar vevs completely Higgs $H$, with no residual discrete or vortex topological sector. Schematically this means we represent $\cM_X$ coupled to the scalar that drives the Higgsing transition as follows:
\be
\label{nonabelianHiggscriterion}
L_{\cM_X+\varphi}=L_{\text{CS}[H,k]}
+\sum_\lambda |D_{R_\lambda}\varphi_\lambda|^2+V(\varphi)
\ee
where 
\be 
\bigcap_\lambda\text{Stab}_H
(\langle\varphi_\lambda\rangle)=\{1\}\,.
\ee
Once such presentation of $\cM_X$ exists, we can then set up the problem within the SymTFT, and further couple to 
 matter $\phi$ on the symmetry boundary, that Higgses $\Neu(G)$ to $\Dir(G)$, which then triggers the flow 
\be
\label{nonabeliananomalousflow}
\fB_{\Neu(G)}^{\cM_X+\phi,\varphi_\lambda}
\ \xrightarrow{\ \langle\phi\rangle,\langle\Phi_\lambda\rangle\ }\ 
\fB_{\Dir(G)} \,.
\ee
The final proliferated phase is as in the general discussion of section \ref{sec:Factorization}
\be
\label{nonabelianframeendpoint}
\cT\ \longrightarrow\ \cN_X=\frac{\cB_X'}{\cE_X}\,,
\qquad
\dim\cN_X=\frac{\dim\cT}{|G|\dim\cB _X}\,.
\ee
Lets carry this out in a specific example where a CS-presentation of $\cM_X$ is available.

\paragraf{Example: $\cT=SU(2)_8$.}
Label the simple lines by their spins $j=0,\frac12,\ldots,4$, and take
$X=(2)$ that has spin $\theta_X=e^{\frac{6\pi i}{5}}$ and is therefore anomalous. The fusion
\be
(2)\otimes(2)=(0)\oplus(1)\oplus(2)\oplus(3)\oplus(4)\,,
\ee
shows that it generates the proper integer-$j$ fusion subcategory
\be
\cB_X=\text{PSU}(2)_8
=\left\{(0),(1),(2),(3),(4)\right\}\subsetneq SU(2)_8\,.
\ee
The simple current $J=(4)$ has $\theta_J=1$ and monodromy
$M_{J,(j)}=(-1)^{2j}$.  It is therefore transparent in $\cB_X$.
Since $\dim\cB_X=\frac12\dim SU(2)_8$, M\"uger's dimension formula gives
\be
\cE_X=\cB_X'=\langle(0),(4)\rangle\cong\Rep(\Z_2)\,,
\qquad \cN_X\cong\Vec\,.
\ee
This is the $\cA_J$ algebra in \ref{sec:SU2kchiral}

Condensing the regular algebra $(0)\oplus(4)$ identifies $(0)$ with $(4)$
and $(1)$ with $(3)$, while the fixed line $(2)$ splits.  If $\tau_1$ and
$\tau_2$ denote the two split lines, the projection functor $F: \cB_X\rightarrow \cM_X$ acts as
\be
\ba
F(0)=F(4)&=1\,,\cr
F(1)=F(3)&=\tau_1\boxtimes\tau_2\,,\cr
F(2)&=\tau_1\oplus\tau_2\,.
\ea
\ee
Here $d_{\tau_i}=(1+\sqrt5)/2$ and
$\theta_{\tau_i}=e^{-4\pi i/5}$, so
\be
\cM_X\cong\overline{(G_2)_1}\boxtimes\overline{(G_2)_1}\,.
\ee
The $\Z_2$ equivariant structure exchanges the two factors.  Consequently
the oppositely oriented auxiliary sector in \eqref{anomalousgraphalgebra} is
$(G_2)_1\boxtimes(G_2)_1$.  Writing the non-trivial line of each factor as
the $\mathbf 7$, the graph algebra is
\be
\ba
\cA_X^{\text{diag}}={}&[(0)\oplus(4)]\boxtimes(1,1)\cr
&\oplus[(1)\oplus(3)]\boxtimes(\mathbf 7,\mathbf 7)\cr
&\oplus(2)\boxtimes
\big[(\mathbf 7,1)\oplus(1,\mathbf 7)\big]\,.
\ea
\ee
Every summand has trivial twist.  Moreover, with
$\varphi=(1+\sqrt5)/2$,
\be
\ba
\dim\cA_X^{\text{diag}}&=10\varphi^2=\dim\cB_X\,,\cr
(\dim\cA_X^{\text{diag}})^2&=\dim SU(2)_8\,\dim\cM_X\,.
\ea
\ee
Thus $\cA_X^{\text{diag}}$ is Lagrangian and
\be
\frac{SU(2)_8\boxtimes(G_2)_1\boxtimes(G_2)_1}{\cA_X^{\text{diag}}}
\cong\Vec\,.
\ee

The same endpoint can be reproduced with a Higgs transition.  On the symmetry boundary
introduce two complex flavors for each $G_2$ factor,
\be
\varphi_{1,A}\in(\mathbf 7,1)\,,\qquad
\varphi_{2,A}\in(1,\mathbf 7)\,,\qquad A=1,2\,,
\ee
together with the $\Z_2$-charged scalar $\phi$.  A single real $\mathbf 7$
leaves an $SU(3)$ stabilizer, and a single complex $\mathbf 7$ generically
leaves $SU(2)$; the flavor multiplicity above is therefore essential.
Generic vevs for the two flavors have trivial common stabilizer in each
$G_2$.  Since $G_2$ is simply connected and has trivial center, no residual
discrete gauge or vortex sector remains.  Locking the two sets of vevs by a
$\Z_2$-invariant potential and condensing $\phi$ gives
\be
SU(2)_8\ \xrightarrow{\ \langle\phi\rangle,
\langle\varphi_{i,A}\rangle\ }\ \Vec\,.
\ee

\section{Conclusions and Outlook}
\label{sec:Summary}

We made a proposal for proliferation transitions $\cT$ to $\cT/\cA$ for any condensable algebra $\cA$ in an MTC $\cT$. 
The key new insight is to utilize the SymTFT to address this question systematically and in generality such that it is applicable to any MTC $\cT$, including non-abelian anyons. 

Not every condensable algebra is a symmetry itself.
Therefore, to make the SymTFT approach viable, the crucial first observation is  to  associate a symmetry to any algebra as follows: consider the generated subcategory, $\langle\cA\rangle$, and its transparent lines 
$\cE_\cA=\cZ_2(\langle\cA\rangle)\cong\Rep (G)$ (M\"uger center). It is the finite
group $G$ of the M\"uger center, that determines the SymTFT to be the (3+1)d gauge theory $\fZ(G)$. The transparent lines then descend from the bulk SymTFT lines, but 
non-transparent lines of $\langle\cA\rangle$ are stuck on the symmetry boundary (which is a non-minimal boundary condition, obtained by stacking and gauging with a non-trivial TQFT). The TQFT that furnishes these non-transparent lines are $\cM_\cA=\langle\cA\rangle/\cE_\cA$. 
The physical boundary is  
generically  also non-minimal with  $\cN_\cA=\langle\cA\rangle'/\cE_\cA$. 

All anyons of $\cA$ then live on $\Bsym$, and the transition occurs now solely on this boundary. 
The order parameter is a multiplet of boundary scalars whose vevs breaks $\Rep(G)$ to the stabilizer $\Rep(H)$ (this is determine by the transparent lines that are in $\cA$, i.e. $\cA\cap\cE_\cA=\Fun (G/H)$), together with a condensation in $\cM_\cA$ by the image of $\cA$ inside it, denoted by $\widehat{\cA}$. The Higgsed phase is then
\be
{\cT\over\cA}={\big(\cM_\cA/\wh\cA\big)\boxtimes\cN_\cA \over H^{(0)}}\,.
\ee
The ``transparency" of  $\langle\cA\rangle$ is what classifies the three types of algebras (i)-(iii) in table 
\ref{tab:types}, and decides how much of the
transition is universal. Type (i) is a $\Rep(G)\to \Rep(H)$ Landau-Ginzburg$^*$ model on the
boundary and needs no further input. 
Type (iii) has an empty SymTFT ($\TwoVec$), $\cT$
factorizes, and the transition is a Chern-Simons matter theory with one scalar per generating anyon. 
Type (ii) is the general case, and splits into a type (i)
Higgsing followed by a strictly smaller instance of the same problem,
$\wh\cA\subset\cM_\cA$. We can rerun the protocol until it   terminates in (i)
or (iii). We illustrated this for abelian $\cT$, for $D(S_3)$, and for $SU(2)_k$
and its double. 

The generality of our approach allows many applications and extensions. 
One obvious omission is the restriction to transparent bosons. 
In general, and as we have seen in the $SU(2)_k\times SU(2)_{-k}$ example, there can be transparent fermions, which makes $\cE_\cA$
super-Tannakian, and requires using a spin SymTFT. It would be interesting to consider this in the future. 
Other examples which would lend themselves very nicely to the analysis in this paper are group quantum doubles $D(G)$ (see \cite{YGSSN}) and metaplectic fusion categories $SO(m)_2$ and their doubles.

\subsection*{Acknowledgments}
\noindent
We are grateful to  
Meng Cheng, Yuhan Gai, Ho-Tat Lam, Ryan Lanzetta, Rajath Radhakrishnan, Chong Wang, Alison Warman, Rui Wen for discussions. In particular we thank Meng Cheng for inspiring us to consider the SymTFT in this context. 
Some of the computations were done and checked with the help of ChatGPT/Codex 5.6 and Claude Code Opus 5 and Fable. 
We thank the KITP for hospitality during the program GenSym25, during which the long journey of this collaboration was initiated. 

The work of AA and SSN is supported by the UKRI Frontier Research Grant, underwriting the ERC Advanced Grant ``Generalized Symmetries in Quantum Field Theory and Quantum Gravity''. The work of Y.Z. is supported by NSFC grant No.12505093 and
the starting funds from University of Chinese Academy of Sciences (UCAS) and from the
Kavli Institute for Theoretical Sciences (KITS).

While we were completing our work the paper appeared \cite{Lu:2026fid} which considers a subset of the  proliferation transitions in $D(S_3)$. Furthermore we were informed of upcoming work on abelian proliferation transitions \cite{MCNS} and non-abelian ones \cite{MCNonab}.

%%%%%%%%%%%%%%%%%%%%%%%%%%%%%%%%%%%%%

\appendix

\section{SymTFTs for 2+1d TQFTs}
\label{app:symtft}

The SymTFT \cite{Ji:2019jhk, Apruzzi:2021nmk, Freed:2022qnc} is a general framework to study symmetry aspects of a theory, separating dynamical and kinematical aspects, by spatially separating them in one dimension higher.
For any 2+1d theory with finite (bosonic) symmetry -- 0-form, 1-form or 2-group or more generally any fusion 2-category -- one can construct the SymTFT in full generality \cite{Bhardwaj:2023ayw, Bhardwaj:2024qiv,Wen:2024qsg,Bhardwaj:2025piv,Bhardwaj:2025jtf}, in terms of a 3+1d $G$ gauge theory $\fZ(G)$, for a finite group $G$, possibly with an anomaly. In the present paper we do not consider any bulk anomalies. 

Here we will be interested in 1-form symmetries that form not necessarily invertible $\Rep (G)$ fusions. 
For $\Rep (G)$ 1-form symmetires the SymTFT is the (3+1)d $G$ gauge theory $\fZ(G)$. Its topological defects form the Drinfeld center 
\be 
\cZ(\TwoVec_G) = \cZ (\TwoRep (G))
\ee
with simple objects the 
\begin{itemize}
\item flux surfaces $\Q_2^{[g]}$, labeled by conjugacy classes, 
\item the Wilson lines $\Q_1^{\bm R}$, labeled by irreps of $G$, 
\item dyons $([g], \bm{R})$, where $\bm{R}$ is an irrep of the centralizer of $g$.
\end{itemize}
On the Dirichlet boundary $\fB_\Dir$ the lines end and the surfaces generate a $G^{(0)}$ 0-form symmetry. The Neumann boundary $\fB_{\Neu (G)}$ gauges it: boundary $G$ gauge fields fluctuate, the Wilson lines restrict to the $\Rep (G)$ 1-form symmetry, and the surfaces end on the non-genuine vortex lines $\beta_1^{[g]}$. This is the symmetry boundary for the 1-form symmetry that forms $\Rep (G)$ (strictly speaking we should refer to this as the fusion 2-category $\TwoRep (G)$).

More generally, the gapped boundary conditions are labeled by a subgroup $H\leq G$ together with an $H$-symmetric TQFT \cite{Bhardwaj:2024qiv,Wen:2024qsg,Bhardwaj:2025piv,Bhardwaj:2025jtf, Bullimore:2024khm, Decoppet:2023rlx}. A subclass gauge all of $G$: stacking a $G$-symmetric TQFT $\cN$ on the Dirichlet boundary and gauging the diagonal $G^{(0)}$ gives the non-minimal Neumann BCs
\be
\label{eq:modified neumann}
\fB_{\Neu (G)}^\cN   := \frac{\fB_{\Dir} \boxtimes \cN}{G^{(0)}} \,.
\ee
The interval compactifications are
\be
\label{N1N2}
\SymTFTS{\fB_{\Neu (G)}^{\cN_1}}{\ \fZ(G)\ }{\fB_{\Neu (G)}^{\cN_2}} = \frac{\cN_1 \boxtimes \cN_2}{G^{(0)}} \,,
\ee
of which \eqref{TSymTFT}, \eqref{Baba}, \eqref{Papa} are the instances used in the main text.

In the Higgs phase of the boundary scalar every Wilson line acquires topological endpoints with multiplicity $\dim R^H$. For the complete Higgsing $H=1$,
\be
\label{wilsontrivialization}
\Q_1^{\bm{R}}\big| \,\cong\, \dim(R)\cdot 1  \,,
\ee
so the $\Rep (G)$ lines act trivially at the endpoint: the 1-form symmetry is not spontaneously broken there, it is gauged away. The vacuum manifold of the Higgsed boundary is the orbit
\be
\label{vacuumorbit}
G\cdot\langle\phi\rangle \,\cong\, G/H \,,
\ee
and the flux surfaces pushed onto the boundary are the domain walls across which the condensate jumps, $\langle\phi\rangle\to\rho(g)\langle\phi\rangle$: they permute the vacua and realize the liberated 0-form symmetry of \eqref{symmetrytrade}. Whether it is spontaneously broken in the compactified theory is decided by the physical boundary: a wall on the symmetry boundary can be slid across the slab and absorbed on $\fB^{\cN}_{\Neu (G)}$, at the cost of acting on $\cN$. For $\cN=\Vec$ the vacua are identified and the endpoint is symmetric -- for $\cT=D(G)$ this is the statement that the Higgs phase of the deconfined $G$ gauge theory is trivially gapped -- while for non-trivial $\cN$ the realization is computed by the $G$ action on $\cN$.

We should note that in our construction, the condensable algebra $\cA$ itself generates only a pre-modular category $\langle\cA\rangle$.
The intrinsic formulation of the bulk in this case is the Walker-Wang theory $\mathrm{WW}(\langle\cA\rangle)$ \cite{Walker:2012mcd}, whose topological defects form $\cZ(\Sigma\langle\cA\rangle)$, the Drinfeld center of the condensation completion of $\langle\cA\rangle$. Only the transparent lines (i.e. the M\"uger center) survive as bulk lines,
\be
\label{WWlines}
\Omega\,\cZ\big(\Sigma\langle\cA\rangle\big)=\cZ_2\big(\langle\cA\rangle\big)=\cE_\cA=\Rep (G)\,,
\ee
see \cite{JohnsonFreydReutter:2024} so the bulk is a (3+1)d $G$ gauge theory, and the non-transparent lines of $\langle\cA\rangle$ are stuck on the symmetry boundary, giving non-minimal stacking with $\cM_\cA$. This however gives then the identification with the topological defects of the (untwisted) $G$ gauge theory,  i.e. the Drinfeld center of $\TwoVec_G$:
\be
\label{WWreduction}
\cZ\big(\Sigma\langle\cA\rangle\big)\,\cong\,\cZ\big(\TwoVec_G\big)\,.
\ee

\section{Proof TQFT Identity for $\cT/\cA$}
\label{app:MasterEq}

In this appendix we will provide a derivation of \eqref{masterendpoint}
\be 
\cT/\cA= \frac{\big(\cM_\cA/\wh\cA\big) \boxtimes \cN_\cA}{H^{(0)}}  \,.
\ee
We carry the proof out in two steps as outlined in \eqref{stagedchain}: 
\be
\cT\ \xrightarrow{\ \Fun(G/H)\ }\ \cT_1= \frac{\cM_\cA\boxtimes\cN_\cA}{H^{(0)}}\ \xrightarrow{\ \ \wh\cA\ \ }\ \cT/\cA\,.
\ee
We should emphasize this is merely a mathematical tool, not a two-step transition.
In the first step we consider the  transparent lines in $\cA$,  $\cA\cap\cE_\cA=\Fun(G/H)$ form a subalgebra of $\cA$, we can use transitivity of condensations, to write 
\be
\label{transitivesplit}
\frac{\cT}{\cA} = \frac{\cT/\Fun(G/H)}{\cA/\Fun(G/H)}\,,
\ee
which is precisely the two steps of \eqref{stagedchain}. For the first step, $\Fun(G/H)$ lies in the $\Rep (G)$ dual to  $G^{(0)}$, and condensing it un-gauges $G$ to $H$,  so that
\be
\label{firststepcompactification}
\cT_1 = \frac{\cT}{\Fun(G/H)} = \frac{\cM_\cA\boxtimes\cN_\cA}{H^{(0)}}\,.
\ee
For the second step we identify the residual algebra. Being a tensor functor, $F$ preserves quantum dimensions and sends the condensable $\cA$ to a separable commutative algebra $F(\cA)$ of the same dimension, which however need no longer be connected. Its number of connected components is
\be
\ba
\dim\, \Hom_{\cM_\cA}\big(1,F(\cA)\big)
&=\dim\, \Hom_{\langle\cA\rangle}\big(\Fun (G),\cA\big)\cr
&=\sum_{R\in\Irr (G)}\dim (R)\,\dim\big(R^H\big)\cr
&=\dim\,\mathrm{Ind}_H^G1 = [G:H]\,,
\ea
\ee
using Frobenius reciprocity in the first line, and in the second that the transparent part of $\cA$ is $\Fun(G/H)=\oplus_R\dim (R^H)\,R$, so that $R\in\Irr (G)$ appears in $\cA$ with multiplicity $\dim (R^H)$.  The residual $G$ action on $\cM_\cA$ permutes these 
$[G:H]$ components transitively with stabilizer $H$: 
they are the $G/H$ degenerate vacua of the first step, and they have the same dimension
\be
\dim\wh\cA= \frac{\dim F(\cA)}{[G:H]}=\frac{\dim\cA}{[G:H]}\,.
\ee
The vev selects the component in the identity coset. This is $\wh\cA$: connected, separable and commutative, hence condensable in $\cM_\cA$, and preserved by the unbroken $H$. Condensing it therefore commutes with the $H^{(0)}$ gauging in \eqref{firststepcompactification}, and we find \eqref{masterendpoint}.

\bigskip

The two extreme cases give rise to type (i) and (iii) algebras:
type (i) has  $\cM_\cA=\Vec$ and $\wh\cA$ is trivial, so \eqref{masterendpoint} reduces to $\cT/\cA=\cN_\cA/H^{(0)}$ with $\cN_\cA=\cT/\Rep (G)$, which is the boundary Higgsing above.

In type (iii), $\langle \cA\rangle$ is modular and we have no boundary gauge group at all (in fact the bulk SymTFT is $\TwoVec$). 
According to Theorem \ref{thm:Muger-gluing} the theory factorizes then 
\be 
\cT=\langle\cA\rangle\boxtimes\langle\cA\rangle'
\ee
and $\wh\cA=\cA$, giving
\be
\label{modularframecondensation}
\cT/\cA \cong \big(\langle\cA\rangle/\cA\big)\boxtimes\langle\cA\rangle'\,,
\ee
so that this becomes anyon condensation only affecting the first factor.

%%%%%%%%%%%%%%%%%%%%%%%%%%%%%%%%%%%%%
\twocolumngrid
\bibliographystyle{ytphys}
\small 
\baselineskip=.7\baselineskip
\let\bbb\bibitem\def\bibitem{\itemsep3.3pt\bbb}
\bibliography{ref}

\end{document}